\documentclass[a4paper,fleqn]{cas-sc}

\usepackage[utf8]{inputenc}
\usepackage[T1]{fontenc}
\usepackage{comment}
\usepackage{float}
\usepackage{tabularx}
\usepackage{microtype}

\usepackage{hyperref} 

\usepackage[numbers]{natbib}

\usepackage[range-units = single]{siunitx}
\usepackage{caption}
\usepackage{subcaption}
\usepackage{lineno}
\modulolinenumbers[1]
\def\tsc#1{\csdef{#1}{\textsc{\lowercase{#1}}\xspace}}
\tsc{WGM}
\tsc{QE}
\let\printorcid\relax 

\begin{document}
\let\WriteBookmarks\relax
\def\floatpagepagefraction{1}
\def\textpagefraction{.001}

\shortauthors{}  
\shorttitle{Demonstrating CBM capabilities by $\Lambda$ Reconstruction with mCBM}   
\title[mode = title]{Demonstrating CBM Capabilities by $\Lambda$ Baryon Reconstruction
in Ni+Ni Collisions with the mCBM Experiment 
at SIS18 of GSI/FAIR}



\author{\textcolor{black}{
A.~Agarwal$^{1}$,             
Z.~Ahammed$^{1}$,             
N.~Ahmad$^{2}$,               
L.J.~Ahrens$^{3}$,            
M.~Al-Turany$^{4}$,           
N.~Alam$^{2}$,                
J.~An$^{4,5}$,                
J.~Andary$^{6}$,              
A.~Andronic$^{7}$,            
H.~Appelsh\"{a}user$^{6,50}$
% ...
}}
\address{$^{1}$Variable Energy Cyclotron Centre (VECC), Kolkata, India}
\address{$^{2}$Department of Physics, Aligarh Muslim University, Aligarh, India}
\address{$^{3}$Justus-Liebig-Universit\"{a}t Gie{\ss}en, Gie{\ss}en, Germany}
\address{$^{4}$GSI Helmholtzzentrum f\"{u}r Schwerionenforschung GmbH (GSI), Darmstadt, Germany}
\address{$^{5}$College of Physical Science and Technology, Central China Normal University (CCNU), Wuhan, China}
\address{$^{6}$Institut f\"{u}r Kernphysik, Goethe-Universit\"{a}t Frankfurt, Frankfurt, Germany}
\address{$^{7}$Institut f\"{u}r Kernphysik, Universit\"{a}t M\"{u}nster, M\"{u}nster, Germany}
% ...
\address{$^{50}$also: Helmholtz Research Academy Hesse for FAIR, Frankfurt, Germany}
% ...

%
















\begin{abstract}
The Compressed Baryonic Matter (CBM) experiment at the upcoming Facility for Antiproton and Ion Research
(FAIR) is a high-rate fixed-target experiment designed to investigate nuclear matter at extreme 
baryon densities in relativistic nucleus-nucleus collisions. To enable high-statistics measurements 
of rare probes, CBM is designed to operate at event rates up to \SI{10}{MHz}. This necessitates 
the development of fast and radiation-tolerant detectors, self-triggered front-end electronics, 
a free-streaming data acquisition architecture, and real-time event reconstruction capabilities.
Prototype versions and pre-series productions of the CBM detector systems have been deployed 
in the mini-CBM demonstrator setup \textit{mCBM} — an experimental precursor comprising 
sub-components of all major CBM systems, installed at the SIS18 facility of GSI/FAIR within the 
FAIR Phase-0 program. 

In 2024, Ni~+~Ni collisions at a kinetic beam energy of \SI{1.93}{AGeV} and an average 
interaction rate of about \SI{250}{kHz} were successfully recorded. This dataset enables a 
detailed evaluation of the operational performance of the detector systems as well as 
the complete CBM data chain, while the reconstruction of rare $\Lambda$ baryons serves 
as a natural benchmark. This paper presents the first results on $\Lambda$ signal reconstruction 
with the mCBM experiment, demonstrating the readiness of the detector technologies and the data chain 
for the upcoming full-scale CBM experiment.
\end{abstract}


\maketitle

\section{Introduction} 
\label{sec:introduction}

The key objective of the Compressed Baryonic Matter (CBM) experiment~\cite{cbm} at the Facility for Antiproton 
and Ion Research (FAIR)~\cite{fair} is to explore the QCD phase diagram at high net-baryon densities and moderate 
temperatures in relativistic nucleus--nucleus collisions~\cite{CbmPhyProg2017}. At the SIS100 facility of FAIR, 
ions with equal neutron and proton numbers can be accelerated up to \SI{14}{AGeV}, heavy ions such as Au 
up to \SI{11}{AGeV}, and protons up to \SI{29}{GeV}. In this energy regime, a rich QCD phase structure may occur, 
including the possible restoration of chiral symmetry and a first-order phase transition. To study rare probes of 
the QCD phase diagram with high precision, CBM is designed to operate at nucleus--nucleus interaction rates 
of up to \SI{10}{MHz}.

To meet these requirements, CBM employs fast, radiation-tolerant detectors, which are read out by a free-streaming 
data acquisition system. Detector data are continuously transported at rates of up to \SI{1}{TB/s} to a large-scale 
computing farm, where real-time reconstruction and an initial stage of online selection are performed.
This triggerless, free-streaming concept poses significant challenges in detector synchronization, data transport, and 
reconstruction, as event candidates must be identified online in the absence of a priori event definition, 
such as those provided by a hardware trigger.

To address these challenges and to commission the CBM detector and data-processing concept under realistic conditions, 
real-size prototypes and pre-series versions of the CBM detector systems are operated in the mini-CBM demonstrator setup
\textit{mCBM}\,--\,an experimental precursor comprising sub-components of all 
CBM systems~\cite{mCBM-proposal2017, mCBM-proposal2020, mCBM-proposal2022}. 
The mCBM experiment is installed at the SIS18 facility of GSI/FAIR and has taken data within 
the \mbox{FAIR Phase-0 program} from 2019 to 2025, enabling operation at interaction rates approaching those foreseen 
for CBM and providing a test bed for detector integration, detector tests, data transport, 
and the data reconstruction framework under realistic experimental conditions.

The primary aim of this paper is to demonstrate the operation of the mCBM experiment as a full-system precursor of the 
CBM experiment, encompassing detector integration, free-streaming data transport, and 
particle reconstruction using the CBM data-acquisition and reconstruction framework. To this end, the reconstruction 
of $\Lambda$ baryons is employed as a benchmark analysis, as it represents a genuine and relatively rare signal 
with a characteristic weak-decay topology, requiring efficient track reconstruction, precise detector calibration 
and alignment, and effective background suppression in a high-rate, triggerless environment.
The use of $\Lambda$ baryons thus provides a realistic validation of the complete CBM data-processing chain, 
demonstrating that non-trivial and rare physics observables can be reconstructed from continuous detector data.
The results presented in this work are based on data recorded in Ni+Ni collisions at a kinetic projectile energy 
of 1.93\,AGeV during the 2024 mCBM campaign.

\section{The experimental setup of mCBM in 2024}    
\label{sec:mcbm_setup}
\begin{figure}[pos=b] 
\centering
\includegraphics[width=0.95\linewidth]{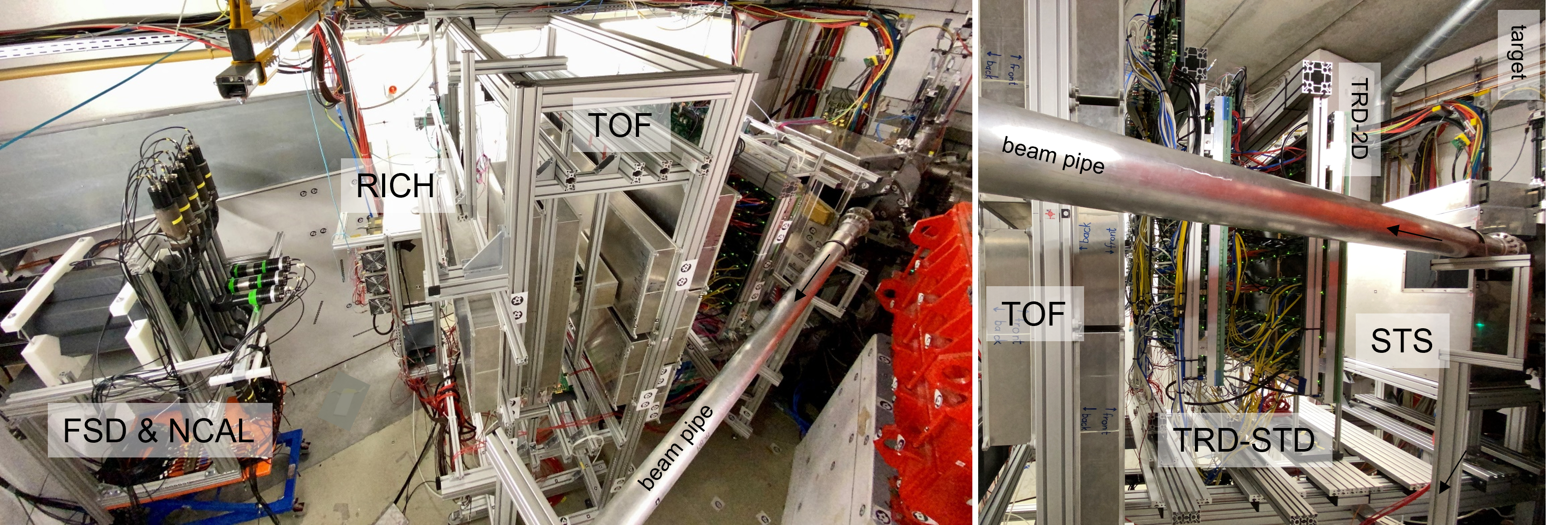}
\includegraphics[width=0.95\linewidth]{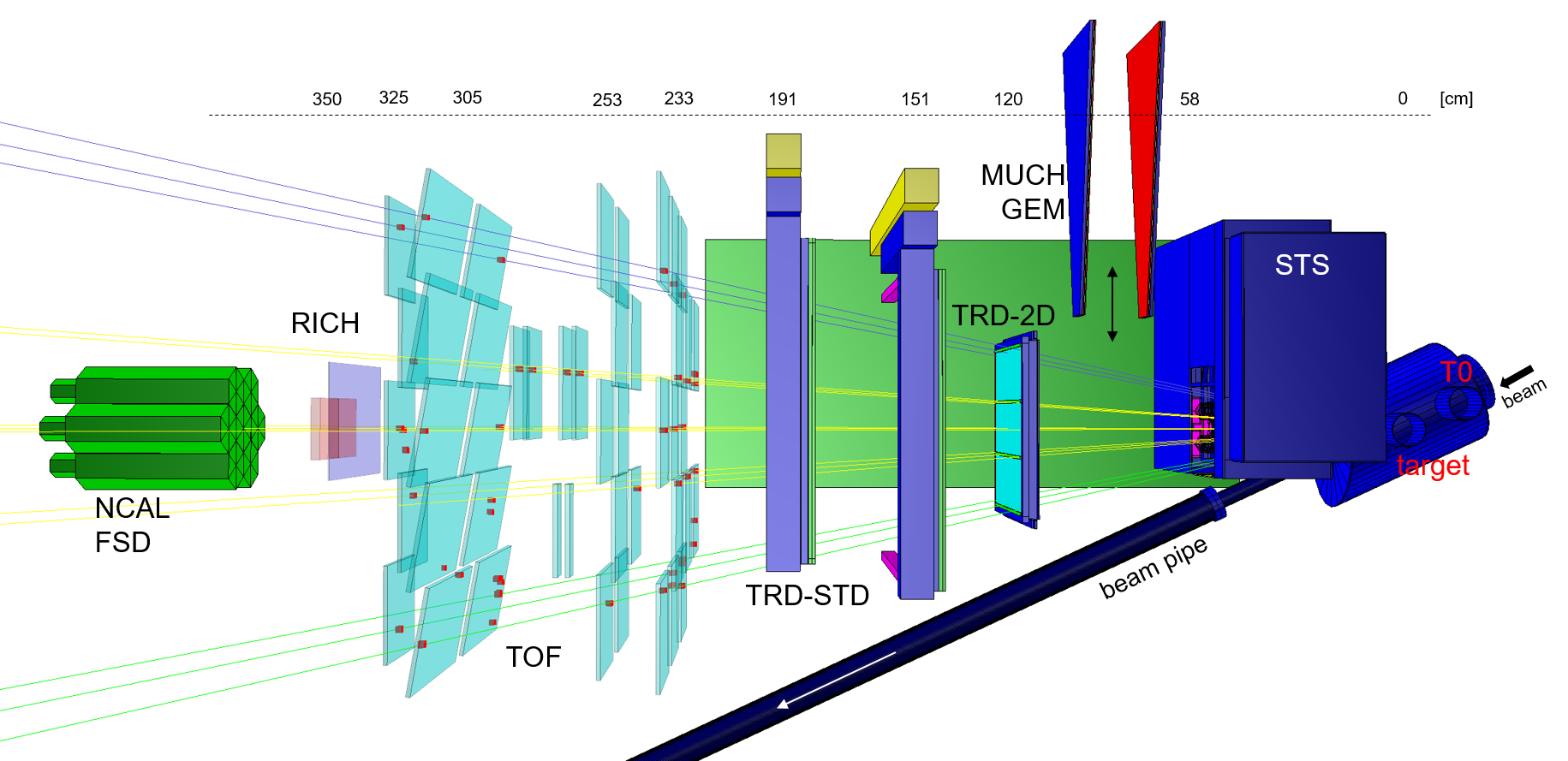}
\caption{Photograph, as of April 2024, and top view ROOT geometry (lower panel) of the 2024 mCBM setup, 
the beam enters from the right.}
\label{fig:mcbm_setup_2024}
\end{figure}

Fig.~\ref{fig:mcbm_setup_2024} shows a photograph of the mCBM 2024 setup as of April 2024 (upper panel) 
and the corresponding ROOT geometry (lower panel). As depicted, the mCBM experiment is positioned downstream 
a solid target under a polar angle of about 25$^{\circ}$ with respect to the primary beam.
A vacuum target chamber hosts a 5-slot target ladder with a remote-controlled step-drive, 
equipped with two Ni and two Au targets, providing 1\% and 10\% interaction probability 
in symmetric reactions as well as an empty-target slot for beam setup and background investigations.
A beam pipe is connected to the vacuum of the target chamber and encloses the primary beam towards 
a beam dump, located about \SI{7}{m} downstream the target at the south end of the experimental area HTD 
of the SIS18 facility, whereas the height of the optical beam axis measures to \SI{2}{m} w.r.t. the floor level. 
Due to the limited space inside the experimental area, mCBM does not comprise 
a magnetic field, and, therefore, measures charged particles produced in nucleus-nucleus collisions 
traversing the detector stations on straight trajectories. 

As visible in Fig.~\ref{fig:mcbm_setup_2024}, the mCBM 2024 setup includes detector stations comprising 
full-size prototypes or pre-series productions of the CBM detector subsystems: 
the Beam Monitoring system (BMON),
the Silicon Tracking System (STS),
the Muon Chamber system (MUCH),
the Transition Radiation Detector system (TRD),
the Time-Of-Flight detector system (TOF),
the Ring Imaging Cherenkov detector system (RICH) and
the Forward Spectator Detector system (FSD) with the Neutron Calorimeter (NCAL).
While the STS, MUCH and TRD were mounted on rail-system sleds of the central beam table, centered  
around a 25$^{\circ}$ polar angle, the TOF and RICH as well as the FSD and NCAL were mounted 
on mobile frames which can be individually positioned for performance studies of the detector systems.  
For data taken during the benchmark run in May 2024 the STS, TRD, TOF, RICH and FSD/NCAL 
were positioned around the 25$^{\circ}$ center line. The individual positions of the detector stations were
measured (pre-aligned) with a precision of about \SI{2}{mm} and included into the ROOT geometry, see lower panel 
of Fig.~\ref{fig:mcbm_setup_2024}. Furthermore, a software-driven alignment procedure determines corrections   
w.r.t. the ROOT geometry (see section~\ref{ssec:alignm}). In the following, the detector subsystems 
are described, with particular emphasis on the BMON, STS, and TOF systems, as they are directly involved 
in obtaining the final result on $\Lambda$ baryons. A brief description is provided for detector systems 
that are not directly used in the present analysis.

\subsection{Detector systems in mCBM}
\label{det}

\paragraph{\textbf{Time-zero detectors of the Beam Monitoring system (BMON)}}
\label{par:bmon}
\noindent
The CBM BMON system consists of a high-speed time-zero (T0) and halo detector~\cite{rost:ibic2023-mop018}.
For the 2024 campaigns, the mCBM setup was equipped with two prototype poly-crystalline 
Chemical Vapor Deposition (pcCVD) diamond detectors to measure the start time (time-zero T0) of the collision. 
A 0.45$\times$0.45\,cm$^2$, segmented into 4 pads, \SI{300}{\micro m} thick pcCVD diamond 
was mounted \SI{45}{cm} upstream the target. 
A 1$\times$1\,cm$^2$, with a metallization arranged in 16 vertical strips, \SI{80}{\micro m} thick 
pcCVD diamond was installed \SI{20}{cm} upstream of the target inside the target chamber. 
Both detectors were read out by TOF electronics (c.f. TOF paragraph below), 
discriminating analog signals with PADI ASICs combined with Get4 ASICs as time-to-digital converter (TDC). 
80\,ps time resolution was achieved for both diamond detectors.

\paragraph{\textbf{The Silicon Tracking System (STS)}} 
\label{par:sts}
\noindent
The design of the CBM STS detector system comprises 876~double-sided double-metal silicon strip sensors 
arranged in eight tracking stations positioned between \SI{30}{cm} and \SI{105.5}{cm}  
downstream of the target inside the gap of the \SI{1}{Tm} superconducting dipole magnet \cite{StsTDR2013}.
Each \SI{320}{\micro m} thick sensor features 1024~strips per side with a strip-pitch of \SI{58}{\micro m}. 
The strips on the p-doped side are inclined by \SI{7.5}{\degree} with respect to the n-side, 
thus providing 2D~tracking information with a single sensor. 
The material budget yields to 0.3~-~1.6\,\%~$X_{0}$ per STS station. 
The eight tracking stations comprise sensors with outer dimensions of $6.2\times\SI{2.2}{cm^2}$, 
$6.2\times\SI{4.2}{cm^2}$, $6.2\times\SI{6.2}{cm^2}$, and $6.2\times\SI{12.4}{cm^2}$.

The STS setup at mCBM 2024 is centered around a polar angle of \SI{25}{\degree} and comprises three tracking stations 
positioned at \SI{17}{cm}, \SI{33}{cm} and \SI{47}{cm} downstream of the target, 
providing active areas of $6\times\SI{6}{cm^2}$ (station STS0), $12\times\SI{12}{cm^2}$ (station STS1) 
and $18\times\SI{18}{cm^2}$ (station STS2), respectively (c.f. Fig.~\ref{fig:lambda-event} in Section~\ref{ssec:reco}). 
The first two stations are built of one and four~(2$\times$2)~~$6.2\times\SI{6.2}{cm^2}$ sensors, 
the third station of five~$6.2\times\SI{6.2}{cm^2}$ and two~$6.2\times\SI{12.4}{cm^2}$ sensors, 
in total 12 sensors across three stations. Each sensor provides 1024~channels 
on each side which are connected by micro-cables with 128-channel ASICs (STS/MUCH-XYTER \cite{KASINSKI2018225}), 
8~ASICs per front-end board (FEB). 
Sensor, micro-cables and ASICs (on a FEB) form a module, which is the functional building block of the STS. 
Up to three modules are mounted on a carbon-fiber ladder. The ASICs of the stations STS1 and STS2 
deploy a single uplink each whereas the module of the first station STS0 provides a five times higher bandwidth 
by operating all five uplinks. Performance studies of the STS, including the operational performance,
time and position resolution, hit reconstruction efficiency, charge distribution, and signal-to-noise ratio 
are discussed in \cite{CBM:2025voh}.

\paragraph{\textbf{The Time-Of-Flight detector system (TOF)}} 
\label{par:tof}
\noindent
The CBM TOF system is the main workhorse for identifying charged particles and is based on Multi-gap Resistive Plate Chambers 
(MRPCs) \cite{FONTE2000201}. An overview of the design of the CBM TOF system can be found in \cite{TofTDR2014}. It consists of 
a large TOF wall of 120\,m$^2$ composed of 230~modules housing about 1400~MRPCs of different granularity (from 4\,cm$^2$ 
for the most forward region to 50\,cm$^2$ at the highest acceptance angles). The flux of charged particles varies between
100\,Hz/cm$^2$ and 50\,kHz/cm$^2$ depending on the scattering angles. At regions above 1.5\,kHz/cm$^2$, the MRPCs are equipped 
with low-resistive glass (see for more information below), while regions below are made of ultra-thin float glass MRPCs. 
The system is designed for a time resolution of 80\,ps at an efficiency higher than 95~\%.

The mCBM 2024 setup consists of a double wall configuration, each consisting of two layers with an active area of 
\SI{152}\,$\times$\,\SI{125}{cm^2} (corresponding to about 3~\% of CBM-TOF). 
All MRPCs were operated with a gas mixture of 97.5\% R134a (C$_2$H$_2$F$_4$) and 2.5~\% Sulfur Hexafluoride (SF$_6$). 
The front and rear walls are separated by a distance of about \SI{80}{cm} (cf. lower part of Fig.\ref{fig:mcbm_setup_2024})
sandwiching several MRPC test counters. However, the test counters are not used in the current analysis and will not be discussed 
in further detail. 

The front wall comprises five horizontally arranged so called M4 modules, each providing an active area of 152\,$\times$\SI{27}{cm^2}. 
Each module houses five MRPCs of type MRPC2 \cite{Deppner:2012zz}. The MRPC2 counters are designed for the intermediate rate region
\cite{Deppner:2012zz} of the CBM TOF wall, requiring a rate capability of up to \SI{5}{kHz/cm^2} charged particle flux. Further
technical details on MRPC2 are given in Table \ref{tab:tech_detais}. MRPCs within M4 modules overlap by \SI{2}{cm} and are 
alternatively staggered, two in front and three in the back (see lower panel of Fig.\ref{fig:mcbm_setup_2024}). The resistive 
electrodes are made of a special low-resistive glass (mentioned above) developed for the high-rate environment 
at CBM \cite{Wang:2001kzg}. The strips are read out from both ends, resulting in a spatial resolution of about \SI{4}{mm} 
along the strip and \SI{3}{mm} across the strips. All readout strips are vertically oriented within the mCBM double-wall setup.

The rear wall consists of three~modules, two of type M0 (see the description below) at the front side and one of type M4 
(see the description above) at the back side, covering the acceptance gap and overlapping by \SI{4}{cm}. 
The so called M0 modules provide an active area of 152\,$\times$\,\SI{53}{cm^2} each,
containing five~MRPC4 detectors (see Table \ref{tab:tech_detais}) with roof-tile arrangement with 10$^{\circ}$ rotation angle 
and \SI{2}{cm} overlap. 

\begin{table*}[ht] 
\centering
\caption{Design parameters of the MRPC detectors, constructed at the Tsinghua University of Beijing (THU) as well as the 
University of Science and Technology of China (USTC).}
\begin{tabularx}{0.7\textwidth}{|X |c|c|} \hline
Technical parameter & MRPC2 & MRPC4 \\ \hline \hline
Produced at & THU & USTC \\ \hline
Active area& 26.8 cm $\times$ 32 cm & 53 cm $\times$ 32 cm \\ \hline
Configuration&double stack&double stack \\ \hline
Number of gaps & 2 $\times$ 4 & 2 $\times$ 5 \\ \hline
Gap size & 0.25 mm & 0.23 mm \\ \hline
Glass type & low resistive & float glass \\ \hline
Bulk resistivity & 10$^{10}~\Omega $cm&10$^{12}~\Omega$cm \\ \hline
Glass thickness & 0.7 mm & 0.28 mm \\ \hline
Number of readout strips & 32 & 32 \\ \hline
Pitch & 1 cm & 1 cm \\ \hline
Active strip length $\times$ width & 26.8 cm $\times$ 7 mm & 53 cm $\times$ 7 mm \\ \hline
\end{tabularx}
\label{tab:tech_detais}
\end{table*}

In total, the TOF double wall configuration of the mCBM 2024 setup possesses 2560 readout channels, with 320~channels per module. 
The readout strips of each MRPC are connected on both sides directly (without cables) to two 32-channel PADI-FEE 
preamplifier-discriminator boards, housing four~PADI-XI ASICs each with 8 channels \cite{Ciobanu:246088}. 
The PADI output signals feature LVDS standard after discrimination, and the initial raw-pulse width is encoded in the pulse length. 
These LVDS signals are routed via twisted pair cables and feed-through PCBs to the exterior of the gas-tight aluminum boxes. 
The trigger-less TDC boards containing eight~GET4 ASICs \cite{Deppe:2009rpa},\cite{TOF:CbmTechNote2026-Get4} (with four channels each) 
convert the pulses into 32-bit words, encoding the channel number, the timing and the time-over-threshold (TOT) information. 
Additional information like slow control messages, synchronization status and error messages are transmitted as well. 
Five of these PCBs are connected to a data concentrator and a transmitter board called TOF-oROB. The TOF-oROB includes 
the radiation hard GBTx ASIC and the VFTx optical transmitter.   

\paragraph{} 
\noindent
Although the following detector subsystems were fully integrated into the various beam campaigns, they are not used 
for the analysis presented here. Therefore, they are not discussed in any detail. 

The instrumented hadron absorber concept of the CBM Muon Chamber (\textbf{MUCH}) system is designed to identify 
muon pairs in high-rate nucleus-nucleus collisions. The 2024 mCBM setup consists of 2~full-size, large area, trapezoidal-shaped 
Gas Electron Multiplier (GEM) modules, designed for station 1 and 2 of the CBM MUCH system \cite{MuchTDR2015}. 
Performance studies in nucleus-nucleus collisions at various collision rates at mCBM are discussed in \cite{MUCH:GEMatMCBM}.
Because of the large material budget, the MUCH system can be moved out of the mCBM acceptance to reduce energy loss 
and multiple scattering, optimizing the experiment setup for the reconstruction of rare $\Lambda$ baryon decays.

The CBM Transition Radiation Detector (\textbf{TRD}) system is based on  Multi-Wire Proportional Counters (MWPCs) 
in combination with an adequate radiator \cite{TrdTDR2018}. To optimize the tracking capabilities, the TRD system implements 
a $x$ and $y$ position sensitivity technology for small polar angles (TRD-2D)~\cite{TrdADD2023} 
and a standard design (TRD-STD)~\cite{TrdTDR2018} for the outer region. 
At the mCBM 2024 setup, a single demonstrator TRD-2D station of size 57\,$\times$\,\SI{57}{cm^2} 
and two (type-5) TRD-STD stations of size 99\,$\times$\,\SI{99}{cm^2} were installed,
centered around a polar angle of \SI{25}{\degree}. 
None of the three stations were equipped with a radiator; instead, they were used as intermediate tracking stations 
to provide position information between the STS and the TOF wall, thereby reducing the number of fake-track combinations.

The CBM Ring Imaging Cherenkov (\textbf{RICH}) detector system uses a \SI{1.7}{m} long gaseous radiator 
(CO$_2$, refractive index $n=1.00045$), focusing mirrors, and Hamamatsu H12700 multi-anode photomultiplier tubes (MAPMTs) 
for efficient detection of Cherenkov photons with high granularity and sensitivity 
down to the near-UV region \cite{RichTDR2013}). Adapted to the smaller collision energies at the SIS18 facility, 
the RICH detector within the mCBM setup is a small-size prototype following the concept of a proximity 
focusing Cherenkov detector with two $20 \times 20 \times 3\,\mathrm{cm}^3$ aerogel radiator blocks ($n=1.05$) \cite{mRICH-Becker2024}, 
mounted on the backside of the TOF mainframe. The produced Cherenkov photons are detected \SI{10}{cm} further downstream 
by 36~MAPMTs, arranged on 2\,$\times$\,3 front-end backplanes, providing a time precision of about \SI{350}{ps}~(RMS). 

The CBM Forward Spectator Detector (\textbf{FSD}) will be used for event plane reconstruction.
A test system was installed downstream the mainframe of the TOF detector system, consisting of two \SI{5}{cm} thick 
scintillating prototype detectors \cite{cbm_pr_23}. At the backside, 
the FSD test system is supplemented by a Neutron Calorimeter (\textbf{NCAL}) array consisting of seven elements 
of \SI{45}{cm} thick, hexagonal shaped plastic scintillator detectors, previously used by the COSY-TOF 
collaboration \cite{cbm_pr_23}.

\subsection{The CBM data transport system under test at mCBM}
\label{daq}

\begin{figure}[pos=b] 
\centering
\includegraphics[width=0.95\linewidth]{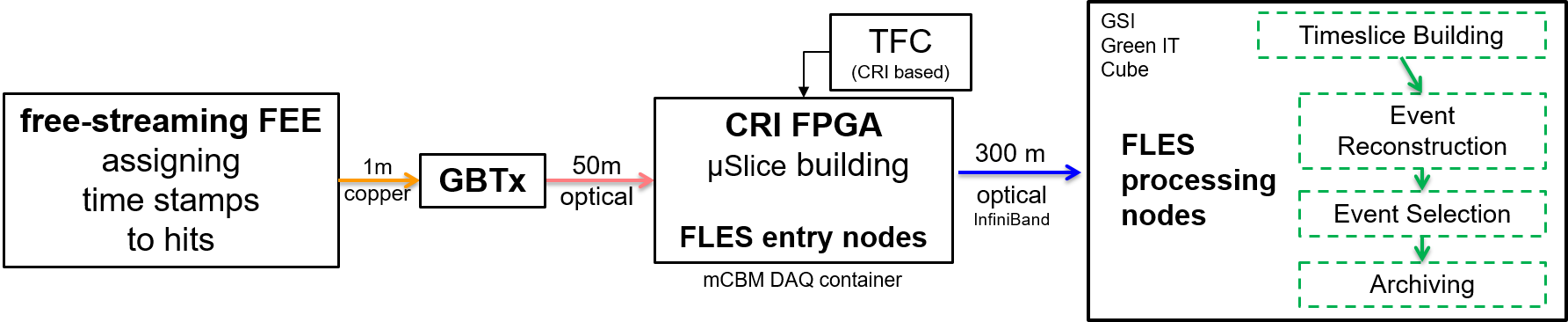}
\caption{Sketch of the CBM data transport chain used as prototype in mCBM.}
\label{fig:DataChain}
\end{figure}

The CBM data transport system~\cite{DaqTDR2022} as it is used in the mCBM experiment 
is sketched in Fig.~\ref{fig:DataChain}.
It includes ultra-fast and radiation-tolerant, subsystem specific ASICs as front-end chips followed 
by CERN GBTx-based radiation-tolerant data aggregation units.
In addition, the RICH, FSD and NCAL detector subsystems use FPGA-TDC based front-end digitization 
instead of dedicated ASICs. The detector front-ends digitize signals 
above threshold and assign time stamps to the hits. By means of a Timing and Fast Control (TFC) system 
the detector front-ends are time-synchronized to the nanosecond level. The data are then forwarded 
via an electrical connection to the GBTx readout board, where the electrical signals acquired 
through a large number of e-links are converted and merged into an optical GBT link, operating 
at a link speed of 4.48 Gbit/s. 

The detector subsystems are interfaced to the PCIe based Common Readout Interface (CRI) boards, 
which are integrated in the Entry Nodes of the First Level Event Selector (FLES). 
The present CRI1 version 
(''BNL-712 v2'') was developed by the Brookhaven National Laboratory (BNL) and is based on the XILINX Kintex UltraScale FPGA.    
The mCBM cave is interfaced to the DAQ container by means of 3~trunk cables, \SI{50}{\m} long, each containing 144~multi-mode optical fibers. 
The entry level stage of the mCBM 2024 FLES consisted of 6~Entry Nodes hosting a total 14~CRI cards, 
which are synchronized with one TFC-Master node. Six ASUS ESC\,8000 G3 GPU-servers were used as Entry Nodes, 
each populated with up to three CRI1 boards. The CRI DAQ hardware and corresponding FPGA design and software versions 
are successfully developed and operated since July 2021.

The long distance connection to the Green-IT-Cube is realized by a \SI{300}{\m} long trunk cable 
holding 144 single-mode optical fibers. A mix of 4~EDR and 2~HDR InfiniBand links offering \SI{800}{Gbps} 
total bandwidth is used to forward the timeslice components from the Entry Nodes to the Build Nodes 
inside the Green-IT-Cube, where 4~Build Nodes and the FLES master server are located. 
These Build Nodes consist of two ASUS ESC~4000~G4 and two ASUS ESC~4000~E10 servers. They assemble the incoming timeslice components 
from various CBM subsystems to timeslices, containing the raw data of all detector subsystems for a specific period of time. 
For mCBM campaigns, a timeslice length of \SI{128}{ms} is used. 

\begin{table}[t]
\centering
\caption{Comparison of the number of readout up-links in mCBM and CBM@SIS100 for the three largest CBM subsystems.}
\begin{tabular}{cccc}
\hline
\textbf{Subsystem} & \textbf{GBT links in mCBM} & \textbf{GBT links in CBM} & \textbf{Ratio} (mCBM/CBM) \\
\hline
STS & 21  & 1728  & 1.2\% \\
TRD & 36  & 1632  & 2.2\% \\
TOF & 16  & 564   & 2.8\% \\
\hline
\end{tabular}
\label{tab:daq_uplinks}
\end{table}

\begin{figure}[pos=!htbp] 
\centering
\includegraphics[width=0.95\linewidth]{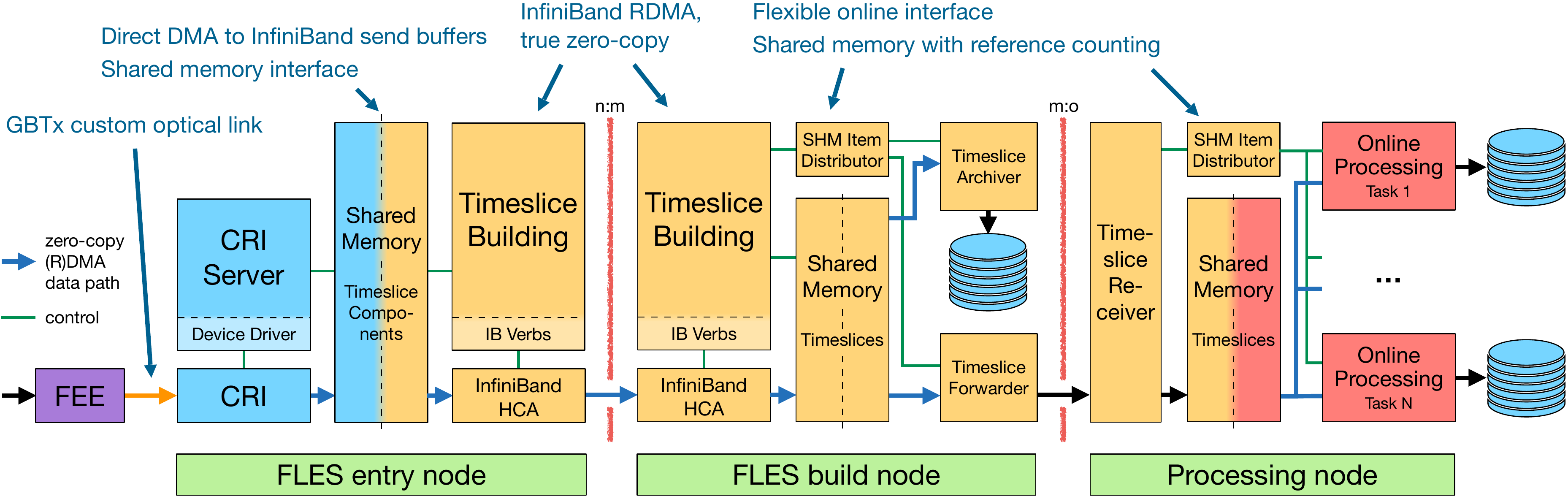}
\caption{The main components of the CBM online data path, adapted from~\cite{refFlesCbm}. The architecture enables zero-copy
data transfer from front-end electronics through entry nodes to processing nodes via a combination of direct memory access (DMA) 
and remote direct memory access (RDMA) techniques. In 2024, data was written to the disk array attached to the FLES Build Nodes.
Forwarding the data to the processing nodes (on the right) was skipped, as the online data processing is still under development.}
\label{fig:datapath_cbm_online}
\end{figure}

The Build Nodes are each equipped with two~14~TB SSDs, three~17~TB HDDs, and three~7~TB HDDs, allowing for a total of 
400~TB local data storage. They allow for a sustained average data rate of 1 GB/s when writing in parallel 
to the solid state drives (SSDs). Since the data is written in parallel, the timeslices are distributed across all disks 
and must all be collected again for the following reconstruction step. During data taking, the data was recorded 
in Time Slice Archive (.tsa) files on this disk array. 
For further processing, the Build Nodes are equipped with a second InfiniBand interface, forwarding the data 
to the processing nodes, which are located in the GSI compute cluster (Virgo cluster) for further analysis and long term storage 
(see Fig.~\ref{fig:datapath_cbm_online}).

Table~\ref{tab:daq_uplinks} compares the number of GBT links present in mCBM with the system size at CBM@SIS100 
for the STS, TRD and TOF subsystems. The sensor of STS station\,0 \cite{CBM:2025voh} was operated 
at a substantial fraction (10\%) of the full anticipated bandwidth for STS sensors at SIS100. Operation close to the limit 
was tested in high-rate tests and will be discussed in a separate publication.

\begin{figure}[pos=!htbp] 
\centering
\includegraphics[width=0.95\linewidth]{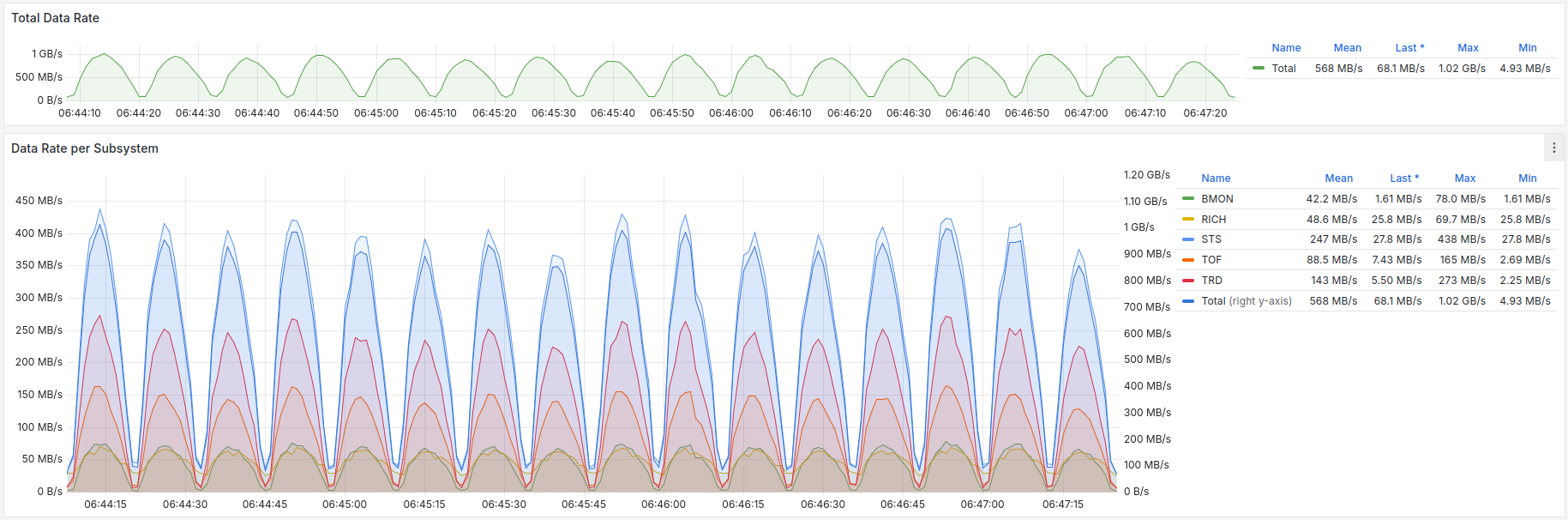}
\caption{Grafana visualisation of data rates in the FLES system for 16 consecutive spills in run 2984, 
taken on May 9th, 2024, around 06:45~am. 
The axis on the left shows the data rate for individual subsystems, while the axis on the right shows the data rate 
for the sum of all subsystems. The amplitude during each spill can be compared with the ‘Max’ column 
in the table on the right, e.g. in the 1st spill ~440 MB/s for STS (light blue) versus 1 GB/s Total data rate (dark blue).}
\label{fig:datarates_09_may_2024}
\end{figure}

Fig.~\ref{fig:datarates_09_may_2024} shows the total data rate in the FLES system, which peaks around \SI{1}{GB/s}, 
for a sequence of 16~spills from run 2984. The individual contribution of the (BMON, STS, RICH, TRD and TOF) subsystems 
to the total data rates is detailed in the bottom part. During spill breaks (no beam) 
the data rate clearly drops down to a local minimum, representing the dark rate of the respective subsystem.

\section{Data Analysis} 
\label{sec:DataAna}
The data analysis chain described here has been implemented in the CbmRoot framework which is derived from FairRoot ~\cite{Al-Turany:2012zfk} 
and so far was only used for simulation studies. Since the primary goal of this study is to demonstrate 
the complete analysis chain up to a final physics result for $\Lambda$ baryons, the data analysis presented here is restricted 
to the three detector systems directly involved in the data evaluation: BMON, STS, and TOF. The analysis is further limited to Ni+Ni collisions, 
for which the yield of $\Lambda$ baryons can be directly compared with published data. A wide range of additional studies 
involving other detector systems can be conducted with the data collected between 2022 and 2025, and these will be addressed in
future publications.

\subsection{Data set} 
\label{ssec:data_set}
The data presented here were collected in 2024 in Ni+Ni collisions with a beam kinetic energy of \SI{1.93}{AGeV}, 
which corresponds to a center-of-mass energy per nucleon of $\sqrt{s_{\rm{NN}}}=2.67$\,GeV. 
The beam delivered approximately $2 \times 10^7$ ions per spill. Each spill lasted for about \SI{8.5}{s} seconds, followed by a break 
of approximately \SI{2.5}{s}. The transverse beam profile had a width of roughly {\color{black}\SI{0.4}{mm}}
as determined from the distribution of the hits on the BMON detector. A nickel (Ni) target with dimensions \SI{15}{mm} $\times$ \SI{35}{mm} 
was mounted on a thin aluminum frame featuring a circular cutout with a diameter of \SI{31}{mm}. The target thickness 
of \SI{4}{mm} corresponds to an interaction probability of about 10\%, resulting in an averaged collision rate 
of approximately \SI{250}{kHz} within the beam spill. 
The dataset\footnote{Run number~2482~-~2485, taken on May 9, 2024, with a total net duration of 5:30\,h of the raw data stream.}
for the results described below was recorded within 5:30\,h in form of timeslice archive files \cite{DaqTDR2022} with 
a total volume of \SI{7.3}{TB}.

\subsection{Event definition and selection}
\label{ssec:EvtBuild}
The event definition is performed by a simple algorithm that can, in principle, also be executed during online data processing. 
The method is based on the smallest entities that can be addressed in the CBM software framework, called digis, 
which represent single time-stamped signals recorded by the free-running front-end electronics channels. Events are built 
on the basis of the multiplicity of pre-calibrated TOF digis within a time interval of \SI{12}{ns}. A coincidence of 4~TOF digis 
stemming from 2~different MRPCs with a BMON T0 hit within a matching \SI{6}{ns} wide time interval is searched for 
in the streamed raw data. When such a trigger signature is found, other TOF and BMON digis found in the time intervals started 
with the fastest TOF digi are added to the event. STS digis are attached to the event in a matching time interval with a width 
of \SI{40}{ns}. 

The time difference of the used subsystem digis relative to the fastest TOF digi mentioned above is shown in Fig. 5. 
For both the time interval selection and the time differences, a pre-calibration is applied to each digi by the addition 
of a channel and/or subsystem specific constant to the measured time. These constants are the same for the whole dataset 
and therefore represent built-in delays in the hardware and firmware design as well as in communication protocols used 
to perform the initial synchronization. 
\begin{figure}[pos=ht] 
\centering
\includegraphics[width=.3\linewidth]{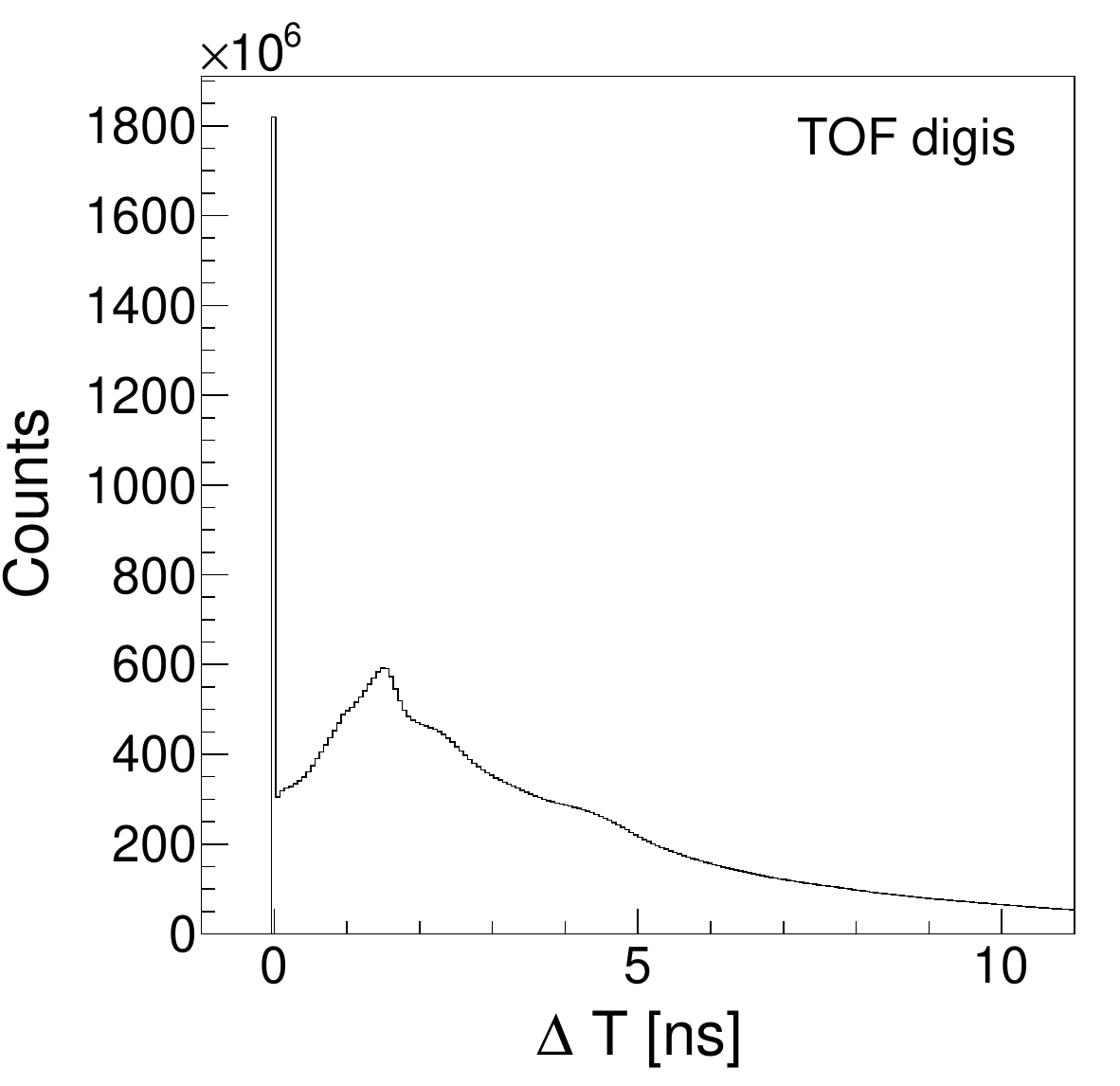} 
\includegraphics[width=.3\linewidth]{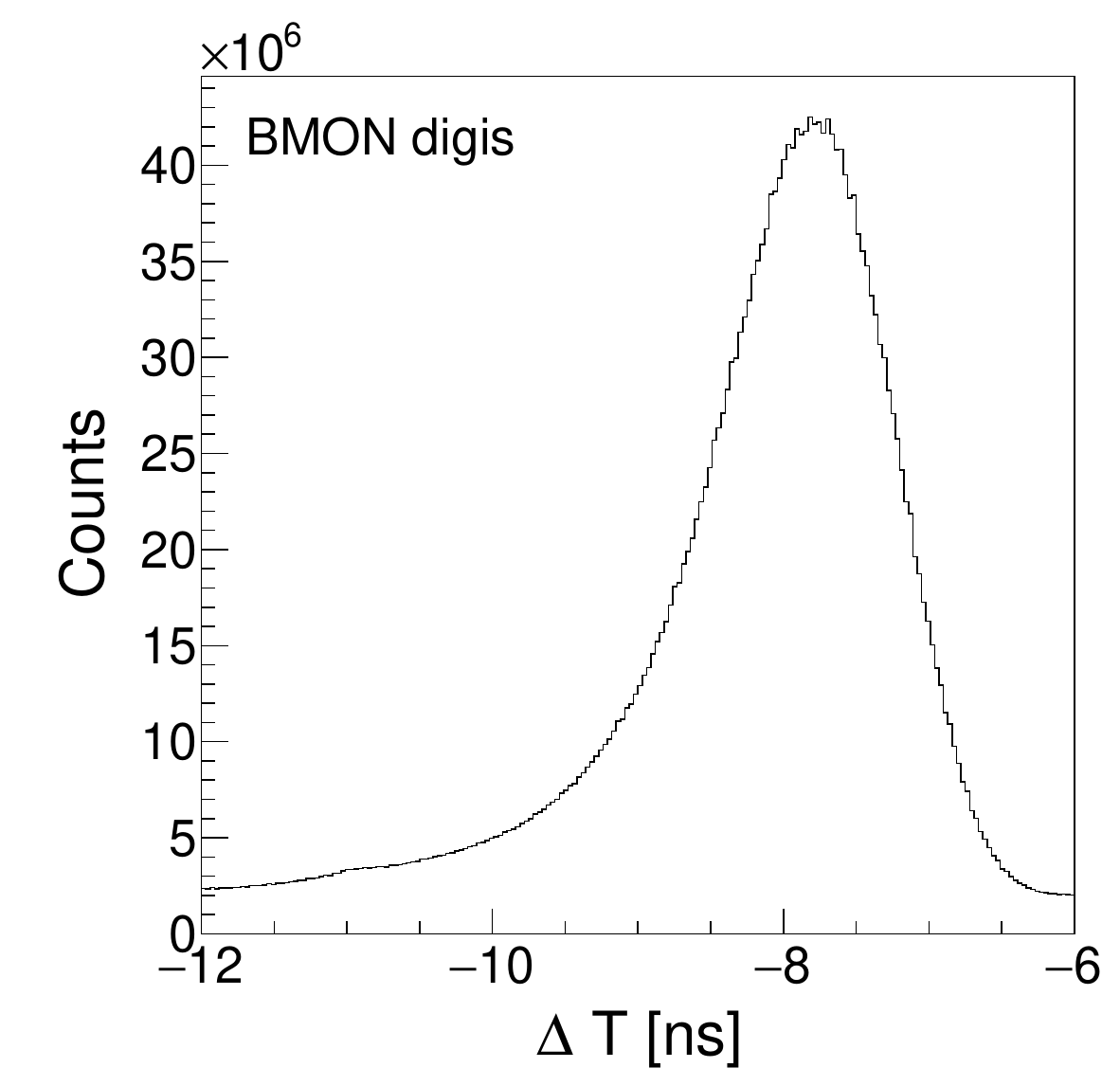} 
\includegraphics[width=.3\linewidth]{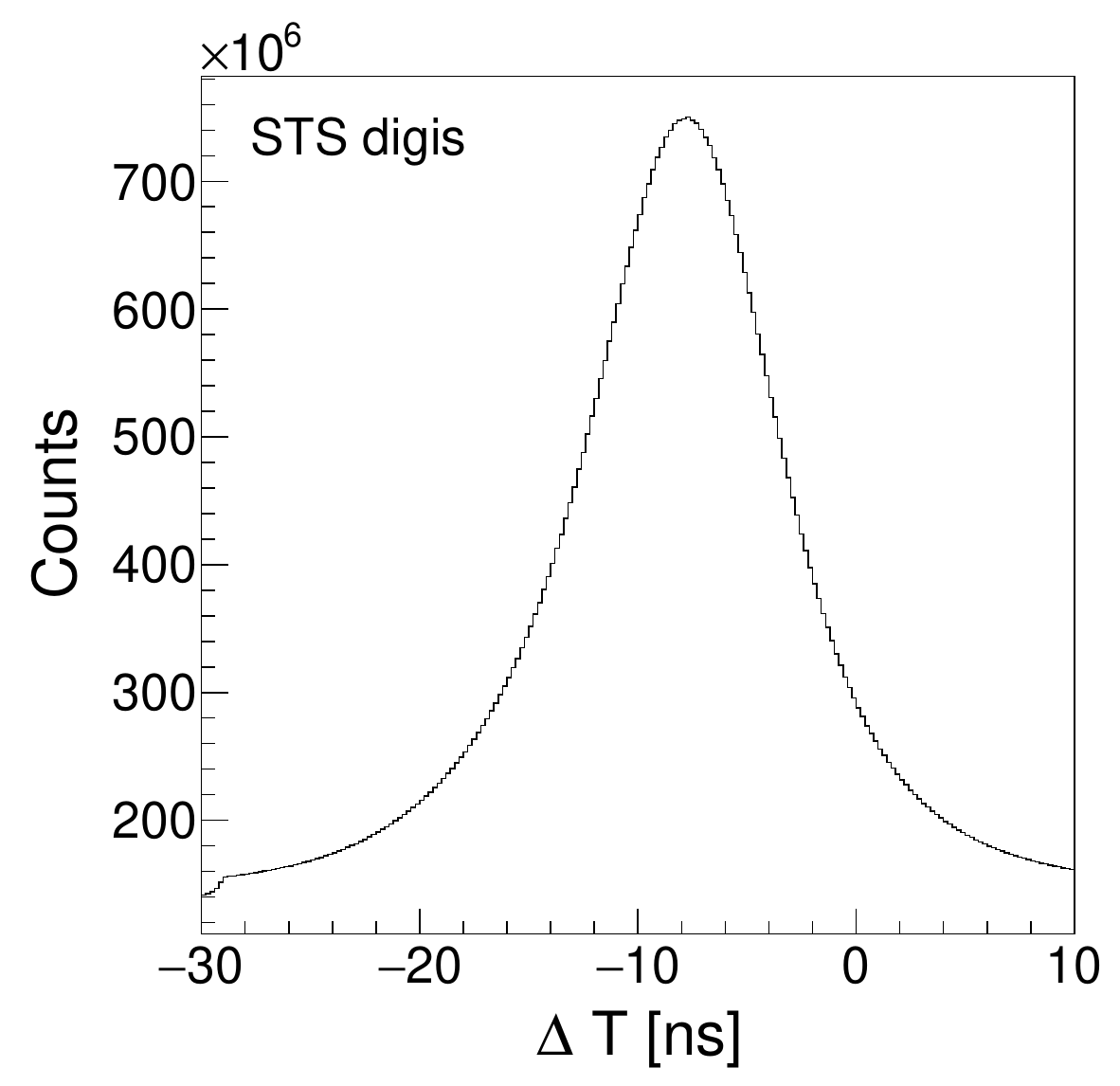} 
\caption{Time difference distribution of digis from the used subsystems to the seed time of the fastest TOF digi in the selected events.}
\label{fig:digiDeltaTseed}
\end{figure}

In total, $N_{\mathrm{event}} = 2.933 \cdot 10^9$ triggered events selected by the software trigger were used for this analysis. 
Relative to the full untriggered data stream, this reduces the data volume to \SI{1.2}{TB}. During the data-taking period, 
$3.56 \cdot 10^{10}$ beam particles passed the target. Assuming an interaction probability of $10\%$, based on 
a geometric (hard-sphere) estimate for Ni+Ni collisions, this corresponds to approximately $3.56 \cdot 10^9$ reactions 
in the untriggered case. The fraction of reactions retained by the software trigger is $\epsilon_{\mathrm{trigger}}\approx 0.82$,
corresponding to a triggered cross section of $\sigma_{\rm{trigger}} = (2.2 \pm 0.2)$\,b. This estimate agrees within systematic
uncertainties with a MC - simulation of the complete setup, where the triggered cross section  
$\sigma_{\mathrm{trigger}}^{\mathrm{MC}}=(2.0 \pm 0.2$)\,b
was found. 
The simulation is based on the Parton--Hadron Quantum Molecular Dynamics (PHQMD) model~\cite{Aichelin:2019tnk}, 
a realistic event generator for nucleus-nucleus collisions at intermediate energies. 
Minimum bias events are generated within an impact parameter distribution $b<\SI{15}{fm}$. The generated events are then processed 
through a full GEANT3-based detector simulation and reconstructed with the same trigger logic as used 
for the experimental data (see Sec.~\ref{sec:sim} for details).

\subsection{TOF Calibration}
\label{ssec:TOFcal}
Since the mCBM setup does not comprise a magnetic field all kinematic quantities have to be derived from time measurements. 
Two subsystems are used for this purpose: the diamond start counters (BMON T0) and the MRPCs of the first TOF wall. The data processing 
for both subsystems is similar, but not identical since the TOF counters feature signals from both ends of the readout strips,
while BMON has only one signal per strip or pad. The calibration proceeds by the following steps.

\begin{figure}[pos=ht] 
\centering
\includegraphics[width=.19\linewidth]{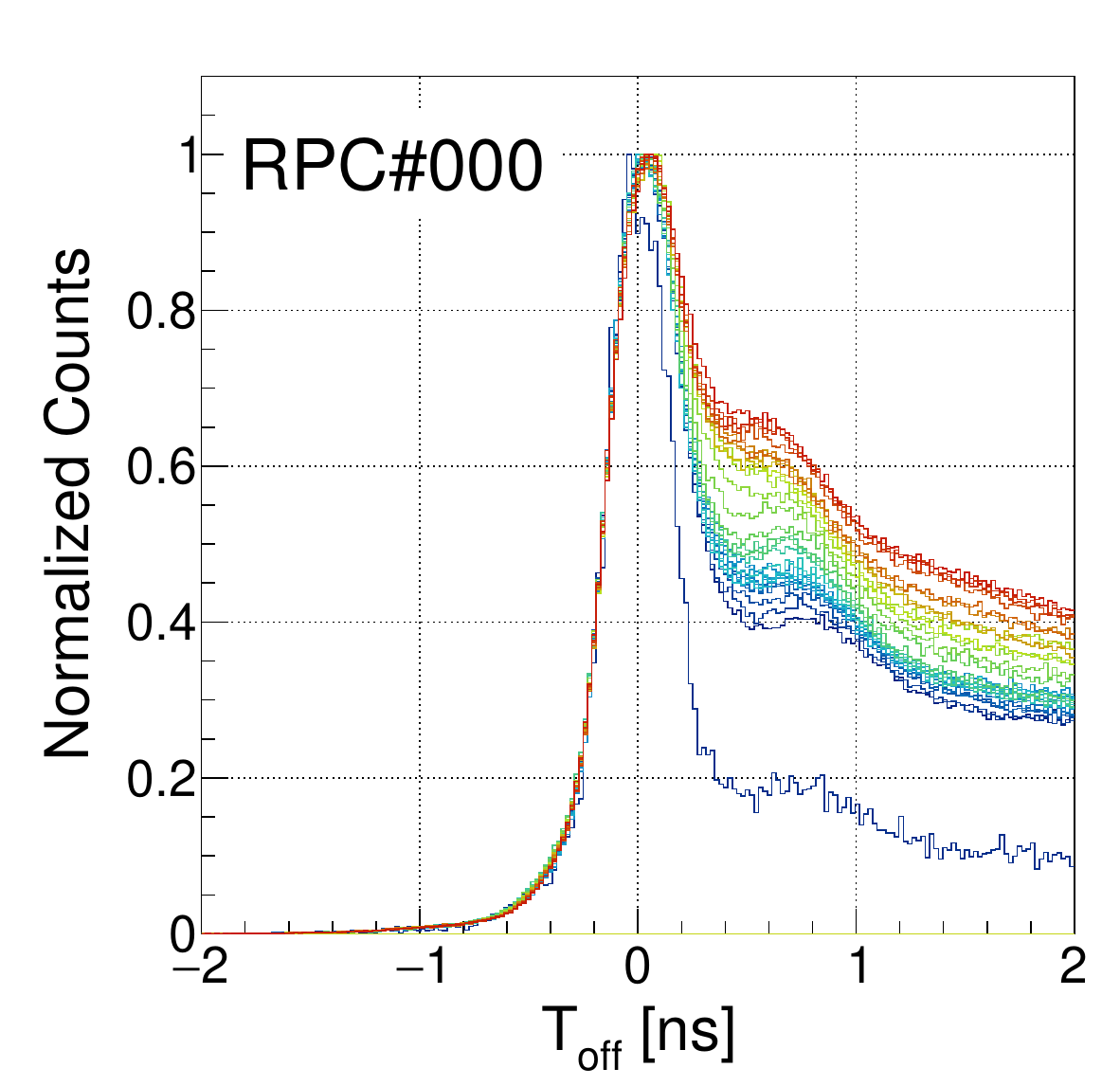} 
\includegraphics[width=.19\linewidth]{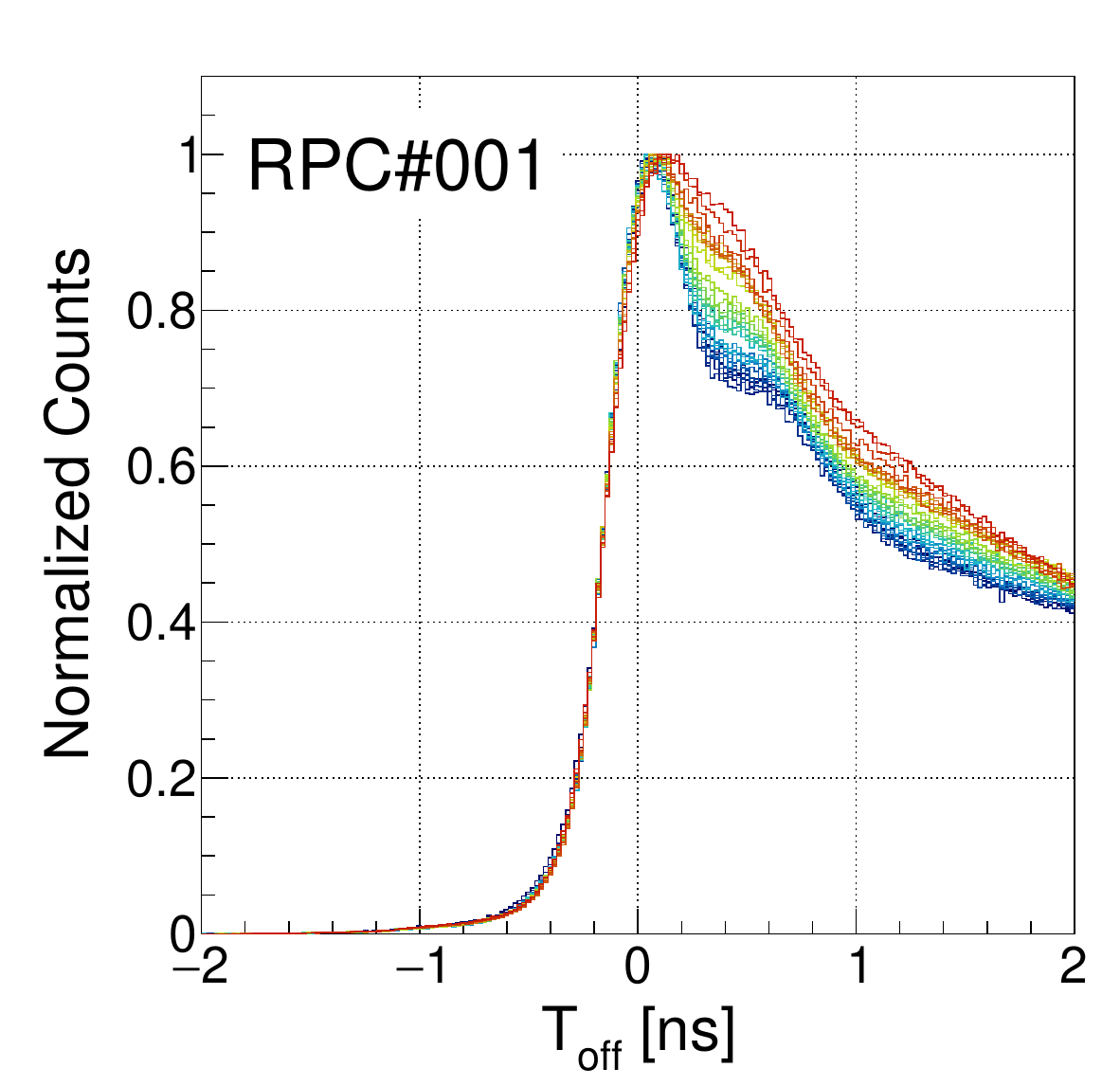} 
\includegraphics[width=.19\linewidth]{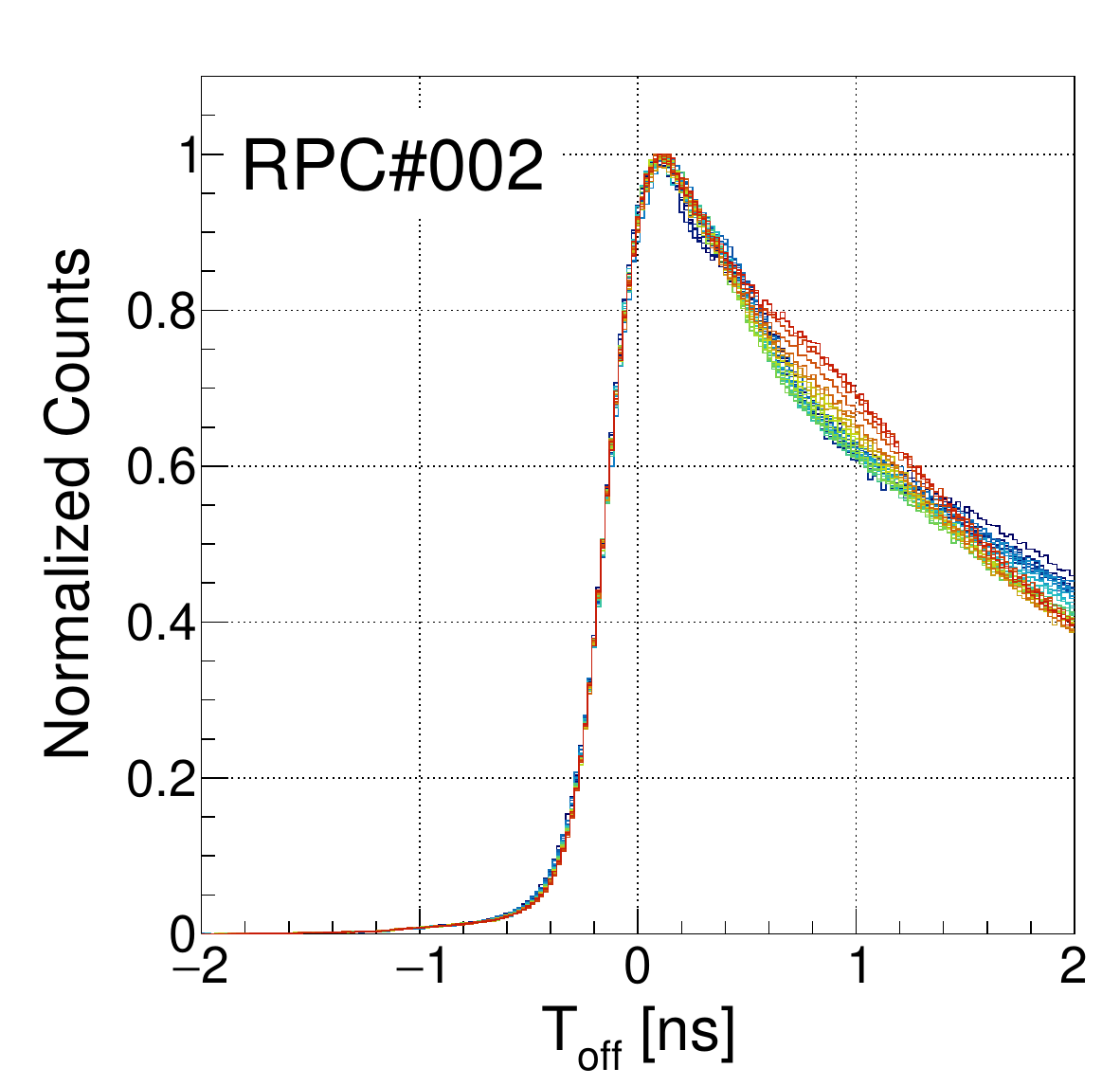} 
\includegraphics[width=.19\linewidth]{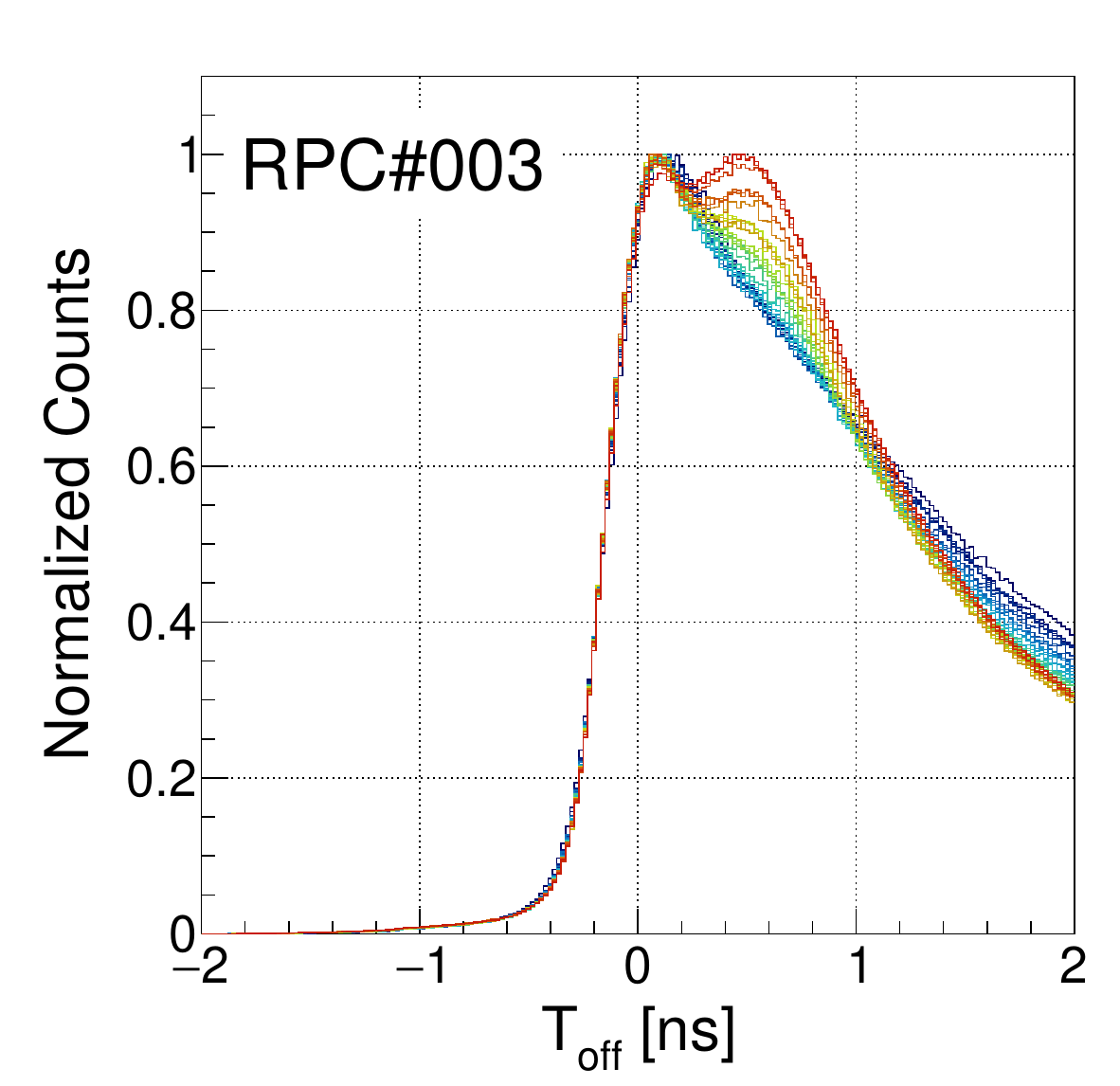} 
\includegraphics[width=.19\linewidth]{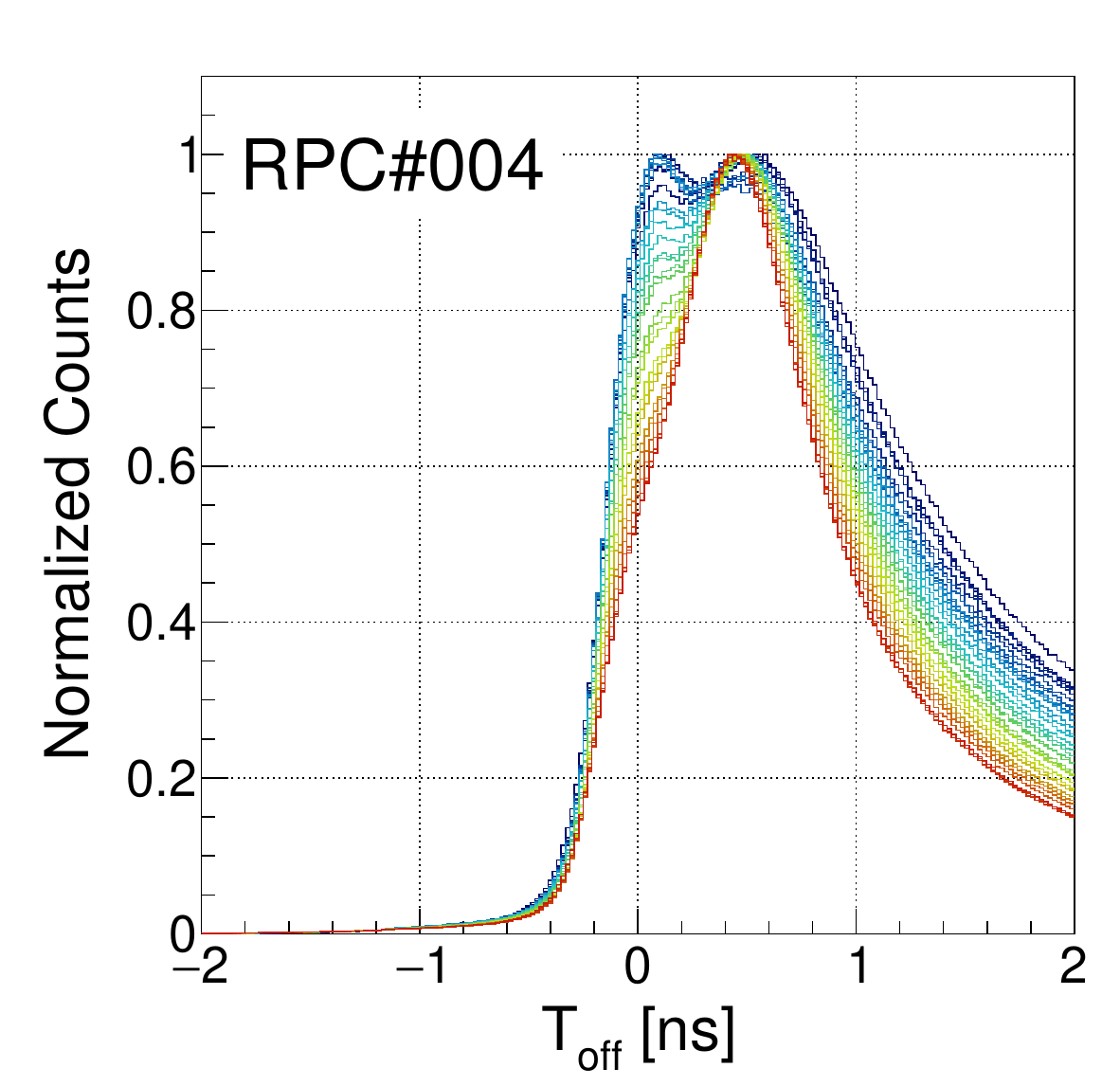} 
\caption{Time difference distributions to the speed of light reference for all strips in the 5 RPCs of the central module of 
the front part of the TOF wall. The strips are indicated by different colours. The spectra are normalized to the peak 
of the distribution.}
\label{fig:TofTdiffc}
\end{figure}

\begin{enumerate}
    \item  The Time-over-Threshold (ToT) values of the digis are used as a measure of the pulse height. To prepare for using them 
    as a weight for the contribution of a single strip in hit reconstruction, the most probable value of the ToT distribution of 
    each channel is scaled by an individual factor to a common reference value. 

    \item TOF hits are formed from two digis originating from 
    both ends of the same readout strip. The arrival time of the hit is given by the average of the arrival times of both digis, 
    while the position along the strips is calculated from the difference of the arrival times. Hits from neighboring strips 
    are merged into one by weighting the contribution of each strip by the corresponding ToT measurement.
    
    \item The time difference distributions of each readout strip are scaled by a counter-specific signal propagation speed 
    to the physical length of the strip. The common offset parameter is adjusted such that both edges of the distribution 
    appear symmetrically around the local origin which is located in the center of the counter.

    \item Since the GET4 discriminators of the TOF system employ a leading edge scheme, the response needs to be corrected 
    for the pulse height dependence (so-called walk correction). 
    The arrival time difference between strips forming a common hit with respect to the common hit's average arrival time 
    is used to determine an additive correction value as function of the ToT value of the involved digis by averaging over many hits 
    with more than one contributing strip. 

    \item The final time-of-flight offsets are determined by inspecting the time difference of the TOF hits to the BMON hits.
    When subtracting the time which is needed by a photon to travel from
    the target to the measured hit position a clear edge is visible 
    in the time-of-flight distributions. These edges are shifted to the origin by an additive constant for each strip. 
    The result of the full procedure is shown in Fig.\,\ref{fig:TofTdiffc}. 
\end{enumerate}

With the described calibration procedure an average MRPC counter time resolution of \SI{90}{ps} is obtained 
by comparing hits in the overlapping regions within the modules. This value is significantly worse than the nominal value 
of better than \SI{60}{ps} that is obtained when using other MRPCs as reference for calibration \cite{Deppner:2016yku}. 
This discrepancy emphasises the need of a good reference, which will be available in CBM by the measured momentum of matched tracks. 
The timing resolution of the BMON counter is found at 79\,ps when both counters are assumed to contribute equally 
to the measured time difference.
    
\subsection{Software-driven detector alignment}
\label{ssec:alignm}
Detector alignment is done by a simple data-driven iterative method starting with measured coarse positions 
entered into a geometry description of the setup in the ROOT framework. The procedure starts with fixing 
the front TOF wall in the laboratory frame. The fine positioning of the STS sensors is achieved by minimizing 
the distance of the measured STS hits to reference TOF hit positions which are obtained by projecting 
the measured TOF hit to the plane of each STS sensor with the assumption that the track stems from the origin 
of the coordinate frame, i.e. in this step it is assumed that all reactions take place at $(0,0,0)$. 
The average deviation of all matched hits in a given STS sensor is used as an initial transverse position shift value 
of the sensor, procedurally implemented by means of root alignment matrices. In the next step, the STS position 
is refined with STS hit pairs or triplets only. 
The final position of each STS sensor in the $x-y$ plane is determined by minimizing the distance ($\Delta x, \Delta y$) 
of the STS hits to the straight line from the nominal vertex to the STS hit pair or triplet which was preselected 
in the first step of the alignment procedure. The $z$-coordinate of each STS sensor is adjusted such that the initially 
observed slope of the deviation of a given hit from the straight line expectation, 
$\Delta x = x_{\rm{measured}} - x_{\rm{line}}$  versus the measured $x$\,-\,coordinate, is minimized. 

\begin{figure}[pos=ht] 
\centering
\includegraphics[width=.65\linewidth]{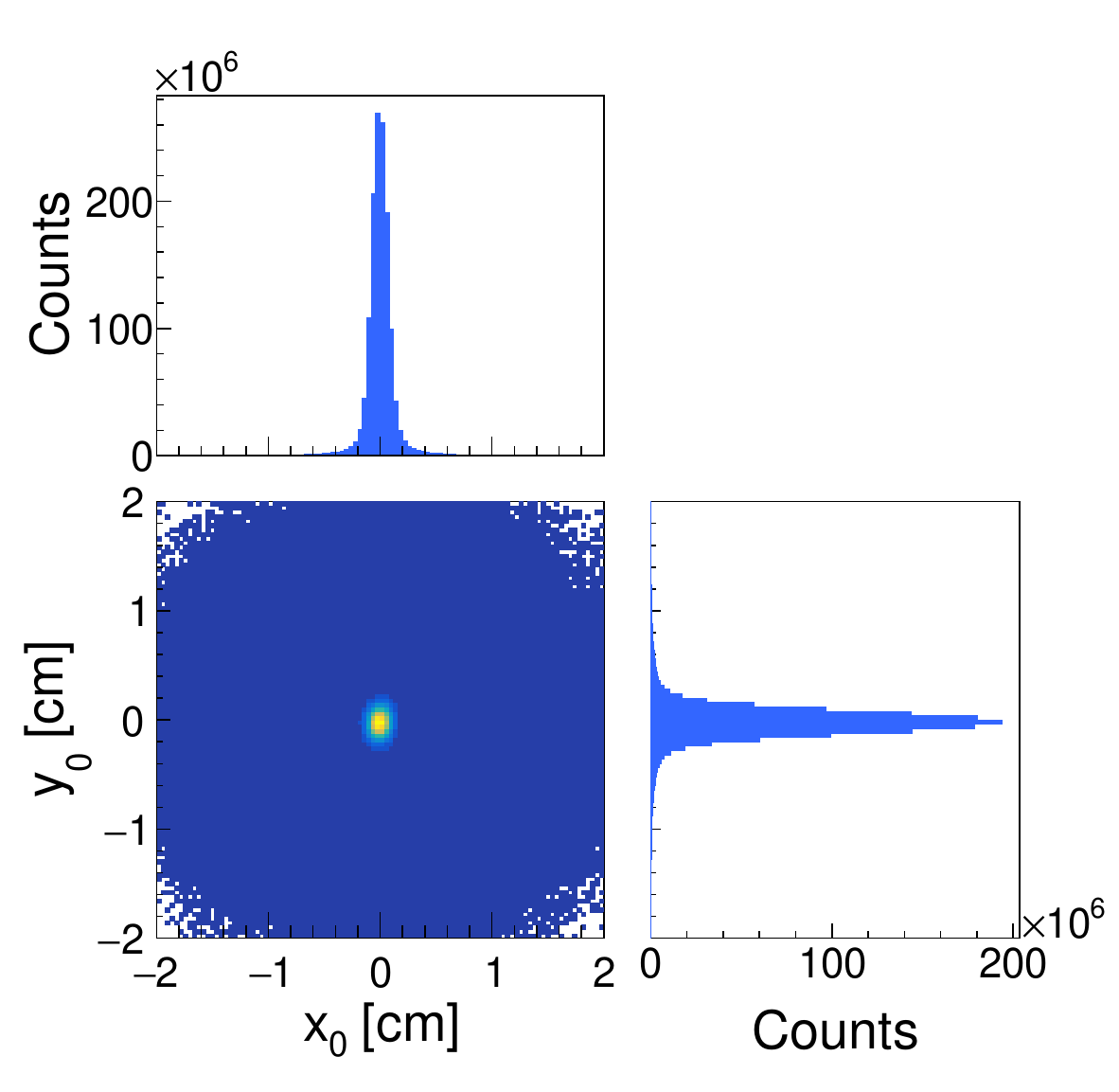} 
\caption{Reconstructed vertex distribution with track candidates used in the alignment procedure.}
\label{fig:vtxAli}
\end{figure}

To visualize the result of the alignment procedure, Fig.\,\ref{fig:vtxAli}
presents the reconstructed intercepts with the $z=0$ - plane of the straight lines 
formed by pairs of the two most forward STS hits of track candidates that are included in the alignment process. 
Fitting Gaussians to the projections onto the axes delivers width values of 
0.074 (0.105) cm in the $x$ ($y$) - directions, respectively.  

\subsection{Track reconstruction, track selection, and particle identification} 
\label{ssec:reco}
\begin{figure}[pos=b] 
\centering
\includegraphics[width=0.7\linewidth]{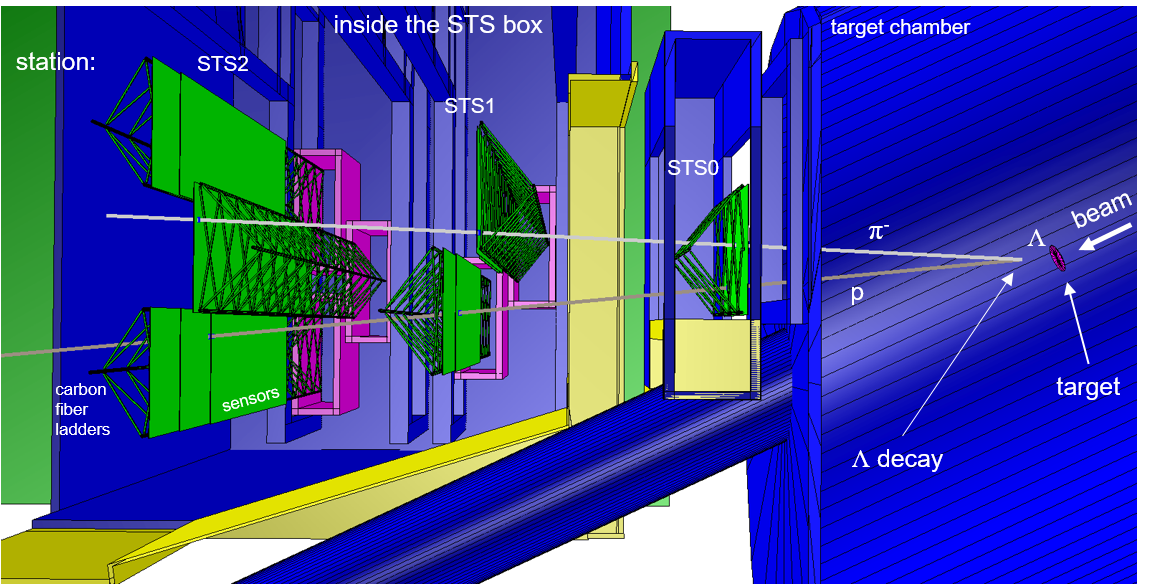} 
\caption{A simulated decay $\Lambda \rightarrow p + \pi^{-}$ is shown within the ROOT geometry 
of the mCBM setup 2024. The Si sensors are visible inside the STS box mounted on carbon fiber ladders 
forming the three stations STS0, STS1 and STS2.}
\label{fig:lambda-event}
\end{figure}

Charged-particle tracks are reconstructed using the Cellular Automaton (CA) algorithm~\cite{refId0}. The CA tracking 
begins by identifying local track segments in the STS and TOF based on geometric proximity and timing compatibility. 
These segments are then linked across the STS and TOF stations to form global track candidates, which are subsequently fitted 
using a Kalman Filter approach that accounts for material interactions. 

$\Lambda$ candidates are reconstructed through the weak decay channel $\Lambda \rightarrow p + \pi^{-}$. 
Fig. \ref{fig:lambda-event} illustrates a simulated $\Lambda$ decay in the mCBM setup. To ensure accurate reconstruction 
of the $\Lambda$ baryon, two key requirements must be met: (i) precise timing information from the TOF detector 
to enable velocity, and consequently momentum measurement, and (ii) good pointing resolution for reliable reconstruction 
of the secondary decay vertex. Therefore, we require each track to have at least one associated TOF hit 
and a minimum of two hits in the STS. To suppress backgrounds from primary protons and pions, we reject tracks 
with a transverse impact parameter less than \SI{0.5}{cm} when extrapolated to the target plane. 

\begin{figure}[pos=hb] 
\centering
\includegraphics[width=.9\linewidth]{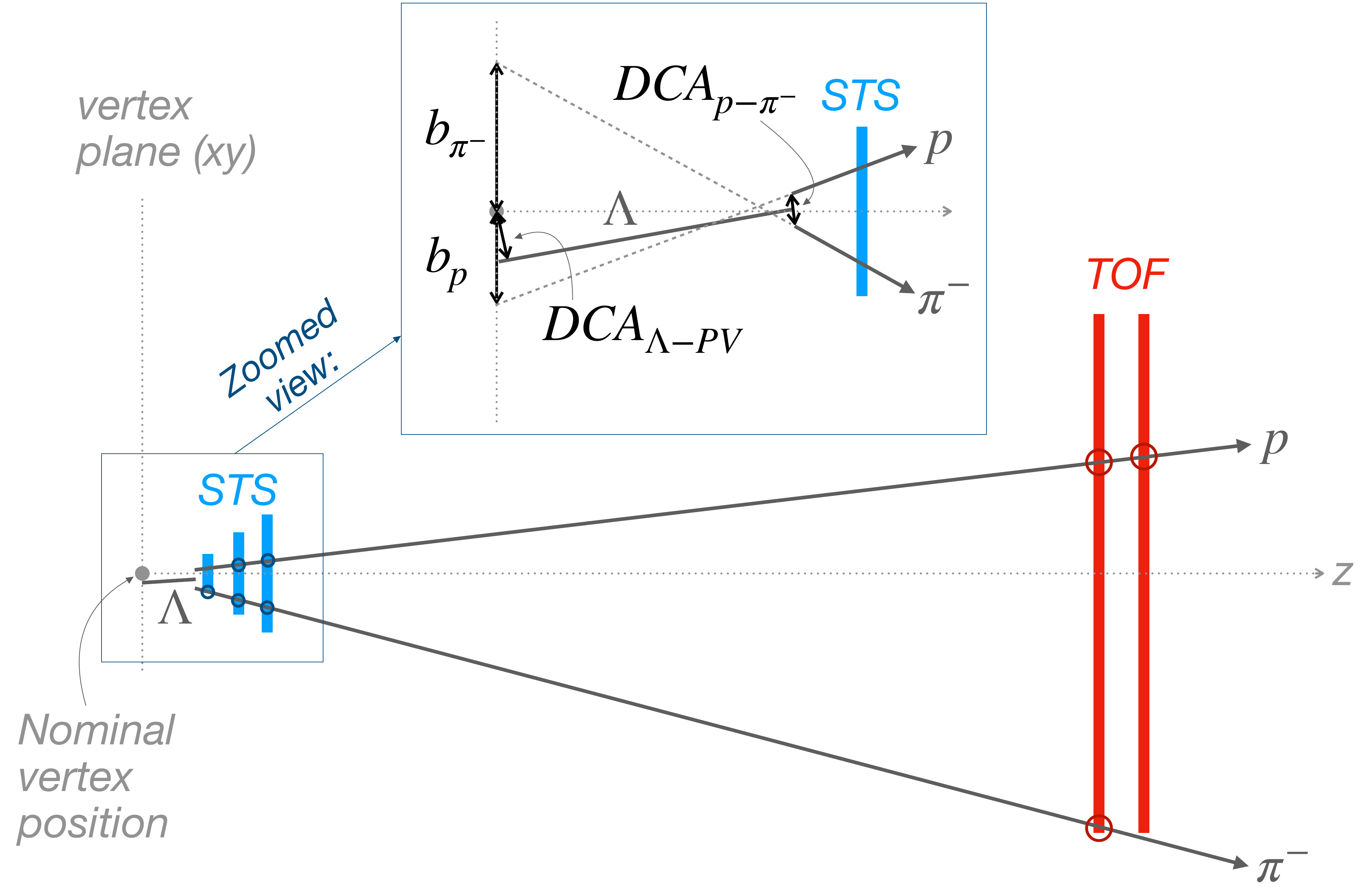} 
\caption{Schematic of $\Lambda$ decay reconstruction in the mCBM setup. The proton track is reconstructed 
using three STS hits and two TOF hits, whereas the pion track is reconstructed using two STS hits and one TOF hit.}
\label{fig:mcbm_cartoon}
\end{figure}

\paragraph{\textbf{Particle identification}} 
Reconstructing $\Lambda$ baryons requires identifying the daughter pions and protons. We exploit the $\Lambda$ decay topology: 
Because the proton has a mass comparable to that of the $\Lambda$, it largely preserves the $\Lambda$’s directionality, 
resulting in a small transverse impact parameter with 
respect to the primary interaction point when the track is extrapolated back to the target plane. In contrast, 
the much lighter pion typically exhibits a larger decay opening angle, leading to a significantly larger transverse impact parameter. 
Thus, the transverse impact parameter evaluated in the target plane is used as the primary 
discriminating observable. In Fig.~\ref{fig:mcbm_cartoon}, the transverse impact parameters are 
labeled as $b_{p}$ and $b_{\pi^{-}}$ for the proton and pion candidates, respectively. 
Monte Carlo studies show that classifying tracks with a transverse impact parameter in the range 
\SIrange{0.5}{1.5}{cm} as proton candidates and those with 
$b > \SI{1.5}{cm}$ as pion candidates optimally preserves $\Lambda$ candidates while maintaining 
a robust signal-to-background ratio. This selection criterion is adopted for the present data 
analysis. In addition, we require 
the pion and proton velocities to lie within \SIrange{25.0}{29.5}{cm/ns} and \SIrange{15.0}{28.0}{cm/ns}, respectively, 
as Monte Carlo studies have shown that these ranges correspond to the expected velocities of $\Lambda$ decay products. 
Table~\ref{tab:pid_criteria} summarizes the particle identification criteria used to classify proton and pion candidates.

\begin{table}[ht]
\centering
\caption{Particle identification criteria used for proton and pion classification.}
\begin{tabular}{lcc}
\hline
\textbf{Particle} & \textbf{transverse impact parameter} $b$ (cm) & \textbf{velocity} $v$ (cm/ns) \\
\hline
Proton ($p$) & $0.5 < b < 1.5$ & $15.0 < v < 28.0$ \\
Pion ($\pi^{-}$) & $b > 1.5$ & $25.0 < v < 29.5$ \\
\hline
\end{tabular}
\label{tab:pid_criteria}
\end{table}

\paragraph{\textbf{Reconstruction of $\Lambda$ candidates}} 
The pion and proton track candidates are used as input to the KFParticle package~\cite{Zyzak2016}, a reconstruction framework for
reconstructing short-lived particles. KFParticle leverages the full covariance matrices of track parameters 
to compute and apply selections based on decay topology variables. The specific selection criteria used in this analysis 
are listed in Table~\ref{tab:topological_cuts}. The $DCA$ variables are evaluated in full three dimensions, 
including those between the $\Lambda$ candidate and the primary vertex and between its daughter tracks, 
as illustrated in Fig.~\ref{fig:mcbm_cartoon}, and are chosen to optimize the signal-to-background ratio. 
The opening angle cut rejects contributions from ghost hits, and the reconstructed $z$-vertex and decay length requirements 
are used to ensure that the $\Lambda$ decays before reaching the first STS station STS0.

\begin{table}[ht]
\centering
\caption{Topological cuts for $\Lambda$ candidate selection}
\label{tab:topological_cuts}
\begin{tabular}{>{\raggedright\arraybackslash}p{6cm}>{\centering\arraybackslash}p{5cm}}
\toprule
\textbf{Topological variable} & \textbf{Selection criteria} \\
\midrule
$DCA_{\Lambda-PV}$ (cm) & $<0.25$ \\
$DCA_{p-\pi^{-}}$ (cm) & $<0.3$ \\
Opening angle between $p$ and $\pi^{-}$ (rad) & $>0.01$ \\
Reconstructed $z$-vertex (cm) & $<17.2$ \\
Decay length (cm) & $5.0-25.0$ \\
\bottomrule
\end{tabular}
\end{table}

In addition to reconstructing $\Lambda$ candidates, KFParticle computes relevant kinematic and geometric variables, 
including invariant mass, momentum, and decay length. However, the velocities obtained from the TOF measurement 
are initially calculated under the assumption that the daughter pion and proton originate from the primary collision vertex. 
In reality, both particles are produced at a displaced secondary vertex from the weak decay of the $\Lambda$ baryon. 
This introduces a slight bias in the estimated flight paths, and consequently leads to small deviations in the reconstructed 
velocities and momenta of the daughter particles.

To account for the true decay geometry, an afterburner module is applied. Starting from an initial estimate 
of the $\Lambda$ four-momentum, the module iteratively determines the $\Lambda$ decay time using its measured decay 
length and velocity. For a given decay time hypothesis, the flight time of each daughter from the secondary vertex 
to its TOF hit is recalculated, allowing updated velocities to be extracted directly from the measured TOF positions and times. 
The corresponding daughter momenta are then recomputed assuming straight-line propagation, which is appropriate 
in the absence of a magnetic field. The updated daughter momenta are then combined to form a new $\Lambda$ four-momentum, 
yielding a refined estimate of the $\Lambda$ velocity and decay time. This procedure is repeated until convergence is achieved. 
The final iteration provides self-consistent daughter kinematics and decay geometry, leading to a robust reconstruction 
of the $\Lambda$ momentum and invariant mass.

\begin{figure}[pos=hb] 
\centering
\includegraphics[width=.60\linewidth]{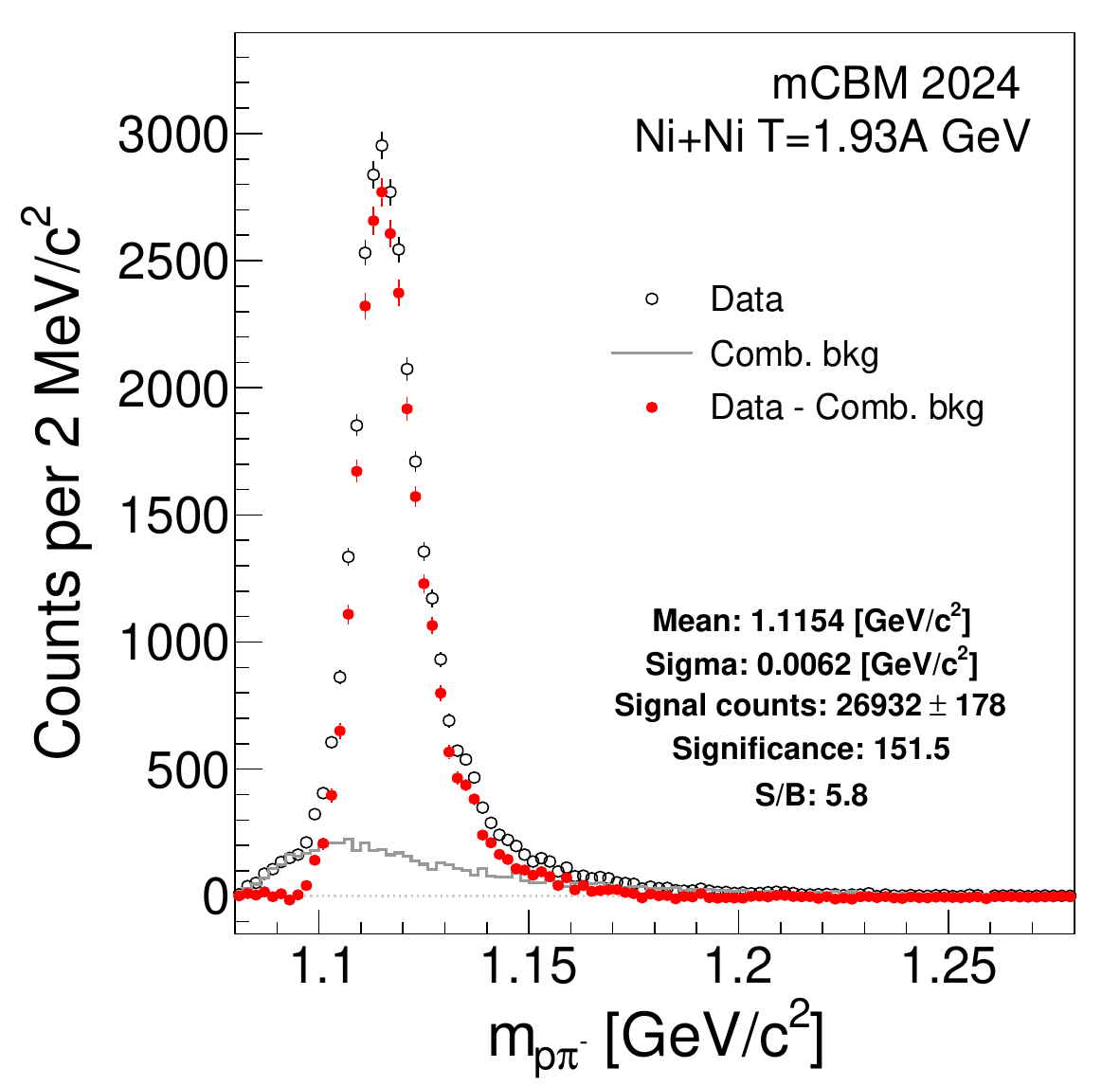} 
\caption{Invariant mass distribution of $p-\pi^{-}$ pairs. Open circles represent the data, the grey histogram represent 
the combinatorial background, and the red dots represent the background-subtracted data. }
\label{fig:lambda_inv_mass}
\end{figure}

\begin{figure}[pos=ht] 
\centering
\includegraphics[width=.6\linewidth]{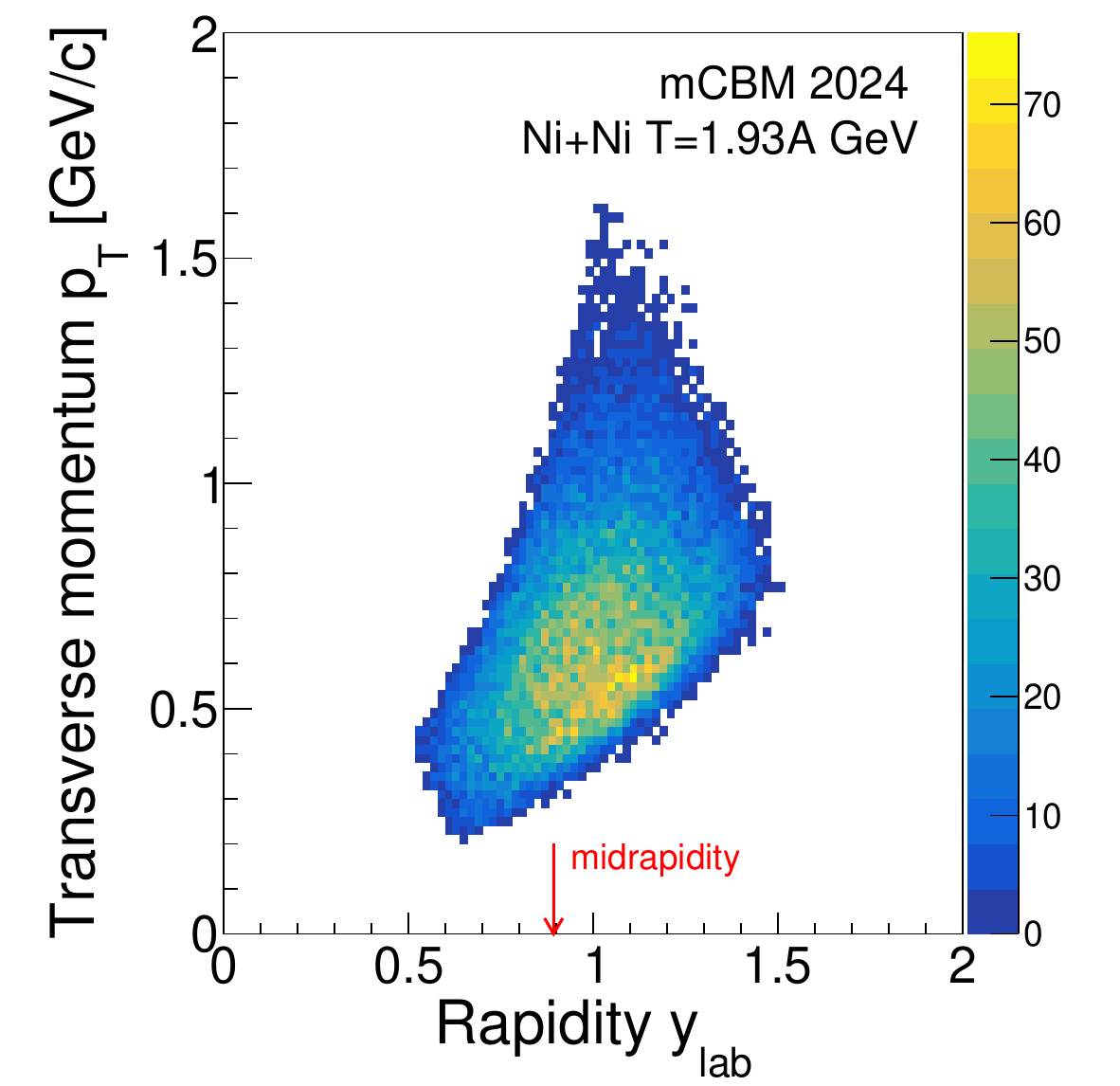} 
\caption{Distribution of the transverse momentum $p_{T}$ vs. rapidity $y_{\mathrm{lab}}$ (in the laboratory frame) 
of reconstructed $\Lambda$ candidates  in Ni+Ni collisions at 1.93 AGeV. The $z$-axis represent the number 
of candidates in each $p_{T}$-$y_{\mathrm{lab}}$ bin. }
\label{fig:lambda_acceptance}
\end{figure}

The invariant mass distribution of the pion–proton pairs is shown in Fig.~\ref{fig:lambda_inv_mass}. 
The $p_{T}$-$y_{\mathrm{lab}}$ distribution of the reconstructed $\Lambda$ candidates passing 
the selection criteria is shown in Fig.~\ref{fig:lambda_acceptance}, where $y_{\mathrm{lab}}$ is the rapidity 
in the lab frame. 
    
\paragraph{\textbf{Background estimation and signal extraction}}
\label{sec:signalextraction}
To estimate the combinatorial background, an event-mixing technique is employed. Proton and pion tracks from 
different events are paired and used as input to the KFParticle reconstruction, with the same selection criteria 
applied as for real-event pairs. The mixed-event background is normalized by matching its integrated yield to that 
of the real data in the invariant mass region \SIrange{1.088}{1.095}{GeV/c^2}.

A pronounced excess over the combinatorial background is observed, peaking near the nominal $\Lambda$ mass. 
To extract the signal counts, the normalized background is subtracted from the real-event distributions, 
and the raw signal yield is extracted using a bin-counting method within a signal window 
around the nominal $\Lambda$ mass, defined as \SIrange{1.095}{1.180}{GeV/c^2}. After subtracting the combinatorial background 
from the data, and interpreting the remaining excess as the $\Lambda$ signal, a signal yield of 
$N_{\Lambda}^{\mathrm{reco}} = 26932 \pm 178$ counts was extracted. 
The corresponding signal significance, calculated as $S/\sqrt{S + B}$, yields  $151.5$, 
and the signal-to-background ratio ($S/B$) is $5.8$.
  
\section{Simulation framework}
\label{sec:sim}
The simulation enables the observed signal to be validated through comparison with experimental data, as well as 
determining the acceptance and efficiency necessary for lifetime and yield measurements.
The simulation chain is implemented within the CbmRoot framework~\cite{Al-Turany:2012zfk} and includes a detailed description 
of the detector geometry and material composition. Monte Carlo events are propagated through the mCBM detector setup 
using the GEANT3~\cite{Brun:1994aa} transport package, which simulates particle interactions with detector materials 
and provides Monte Carlo points (MCPoints) as output. These MCPoints serve as input for the digitization step, 
which simulates the detector response, including the modeling of signal formation in each sub-detector 
(e.g. strips in STS sensors), electronic noise, detector inefficiencies, and timing response. 

The digitized signals (digis) are then fed into the standard CBM reconstruction chain, where the digis are converted 
into clusters and spatial hits. Track reconstruction is subsequently performed using Cellular Automaton seeding followed 
by Kalman Filter–based track fitting. Finally, the reconstructed tracks are used for $\Lambda$ candidate reconstruction. 
Additionally, the beam spot width in the transverse plane is fixed at \SI{0.07}{cm}, which is close to the width of 
the beam profile observed during the experiment. The effective time resolution for the simulations is set to \SI{120}{ps}, 
representing the combined performance of the TOF and BMON.

Dead channels in the STS are identified from experimental data and masked in the simulation. 
Furthermore, certain regions of STS2, while not completely non-functional, exhibit reduced efficiency. 
To reproduce this behavior in the simulation, STS hits in STS2 are randomly removed with a probability of 20\%. 
In addition to these localized inefficiencies, the overall hit detection efficiency of the STS stations is quantified 
using $\Lambda$ daughter tracks originating from a vertex located in front of the first STS station. Combining this 
with the TOF hit detection efficiency provides access to the probability of detecting both daughter tracks, $p$ and $\pi^{-}$. 
The corresponding combined detection efficiency is determined to be $0.86 \pm 0.10$ and is incorporated into the detector response 
in the simulation.

Finally, the efficiency loss due to residual detector misalignment in the data is studied using dedicated MC simulations. 
Since the nominal simulation does not include residual misalignment effects, a pre-misaligned detector geometry is introduced 
at the simulation level, and the same alignment procedure as used for the data is applied. 
The resulting efficiency reduction is characterized by a factor of 
$0.84 \pm 0.10$. Together, these effects provide a realistic description of the detector performance in the simulation, 
enabling a consistent comparison between simulated and reconstructed quantities for efficiency determination 
and signal extraction.

\subsection{$\Lambda$ source parametrization}
\label{ssec:lambdaparam}
The input sample distribution of the $\Lambda$ was tuned using published results of the FOPI Collaboration~\cite{FOPI:2007usx}. 
Under the assumption of a thermal Boltzmann-like source, the transverse momentum $p_{\mathrm{T}}$ spectrum is described by
\begin{equation}
\frac{\mathrm{d}^{2}N}{\mathrm{d}p_{\mathrm{T}}\,\mathrm{d}y_{\mathrm{lab}}}
\sim
p_{\mathrm{T}} \cdot \sqrt{p_{\mathrm{T}}^2 + m_0^2} \cdot
\exp \left( -\frac{\sqrt{p_{\mathrm{T}}^2 + m_0^2}}{T_{\mathrm{B}}} \right) \cdot
\exp \left( -\frac{(y_{\mathrm{lab}} - \mu_y)^2}{2\sigma_y^2} \right),
\end{equation}

where $T_{\mathrm{B}} = T_{\mathrm{eff}} / \cosh(y_{\mathrm{c.m.}})$ is the Boltzmann temperature,
$m_0$ is the $\Lambda$ mass, and $y_{\mathrm{lab}}$ denotes the rapidity in the laboratory frame.
The rapidity distribution is described by a Gaussian with mean $\mu_y = 0.893$, corresponding to the center-of-mass rapidity 
in the laboratory frame for the fixed-target configuration. The width $\sigma_y$ reflects the longitudinal spread of the source. 
The FOPI measurements in central Ni\,+\,Ni collisions at the same collision energy indicate that the source 
can be parameterized by  
\begin{equation*}
T_{\mathrm{eff}}^{\mathrm{FOPI}} = 119 \pm 1~\text{(stat.)}^{+9}_{-7}~\text{(syst.)} \rm{ MeV}, 
\quad 
\sigma_y = 0.386 \pm 0.009~\text{(stat.)}^{+0.047}_{-0.031}~\text{(syst.)}.
\end{equation*}

\noindent 
As centrality cannot be determined with sufficient resolution in the mCBM data, the PHQMD model was used 
to estimate the difference in the Boltzmann temperature parameter between the FOPI centrality selection and 
the software-triggered event sample used in the mCBM analysis. The PHQMD was chosen because it has been verified 
to accurately reproduce the centrality dependence of spectra at a comparable collision energy of 
$\sqrt{s_{\rm{NN}}}=3$\,GeV~\cite{Zhou:2025zgn}.
By fitting the PHQMD $\Lambda$ spectra with a Boltzmann function, the extracted $T_{\mathrm{eff}}$ for the software-triggered 
event sample is found to be approximately $8\%$ lower than that for central collisions, due to enhanced radial flow 
in central collisions. Accordingly, scaling down the FOPI results by $8\%$, $T_{\mathrm{eff}}~=~110~\mathrm{MeV}$ 
and $\sigma_y = 0.386$ are used  as $\Lambda$ source parameters. The azimuthal angle $\phi$ was generated uniformly 
over the interval $[0, 2\pi]$.

\begin{figure}[pos=t]
\centering
\includegraphics[width=0.6\linewidth]{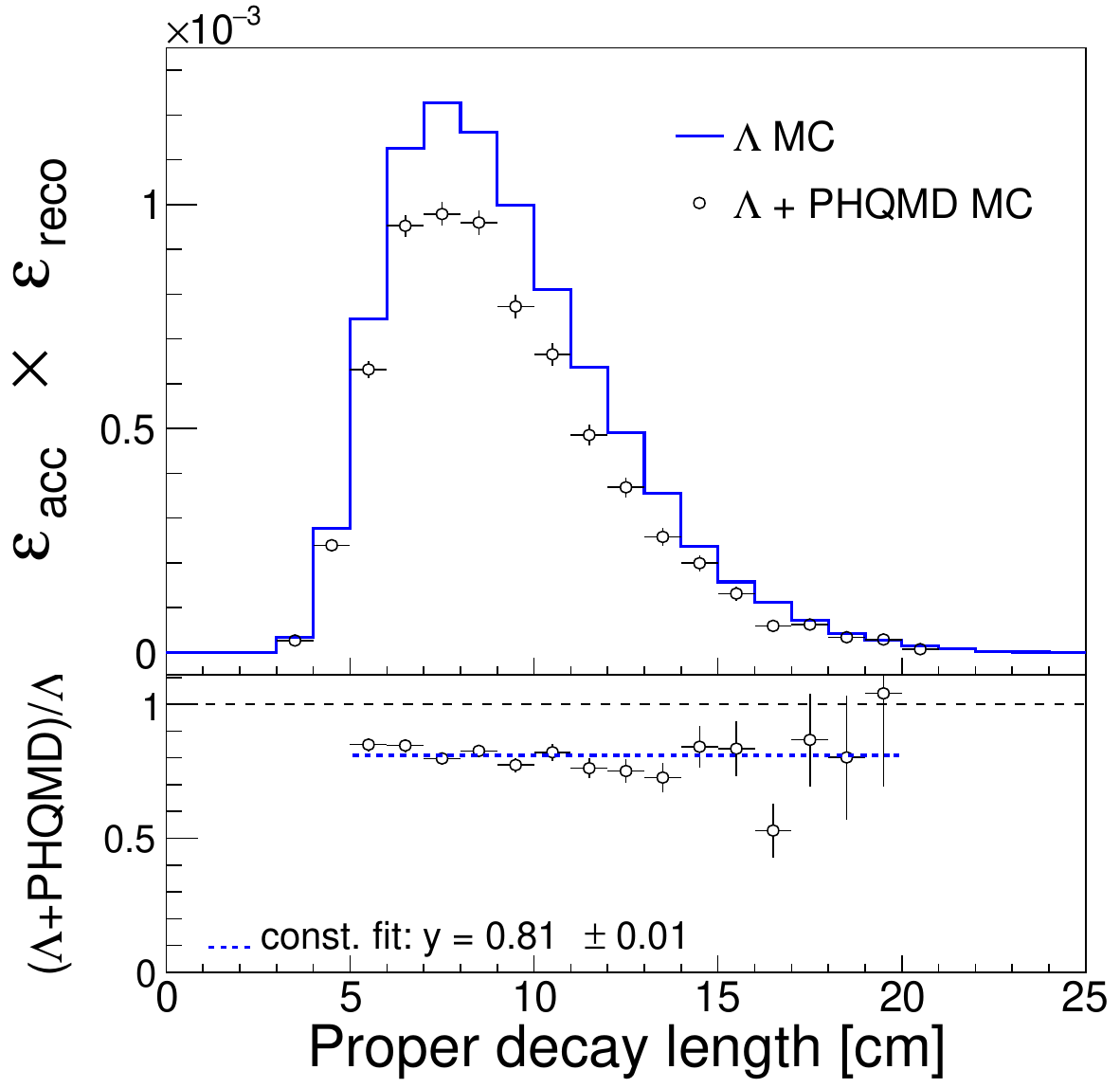}
\caption{ Acceptance $\times$ reconstruction efficiency of reconstructed $\Lambda$ candidates from single-$\Lambda$ (blue histograms) 
and $\Lambda$ + PHQMD embedding (open circles) as a function of the proper decay length. The ratio of the two is shown 
in the bottom panel, along with a constant fit.}
\label{fig:mceff}
\end{figure}

\subsection{Track embedding} 
\label{ssec:phqmd}
To assess the reconstruction efficiency and signal extraction performance, the track embedding technique was used.
In this approach, simulated signal tracks from $\Lambda\rightarrow p + \pi^{-}$ decays are embedded into 
PHQMD events. It has been demonstrated that the PHQMD model reproduces charged particle multiplicities sufficiently well 
at comparable collision energies~\cite{Kireyeu:2024hjo, Zhou:2025zgn}. Furthermore, the STS and TOF hit multiplicities
observed in mCBM data, when selecting events containing a $\Lambda$ candidate, match with simulations using central PHQMD events 
as input. Hence, tracks from a single $\Lambda$ baryon decay are embedded into a central PHQMD event with 
an impact parameter $b < 1.5$\,fm. The combined data are then passed through the standard reconstruction chain, 
allowing a realistic evaluation of the signal reconstruction performance.

\subsection{Acceptance and efficiencies} 
\label{sec:eff}
The upper panel of Fig.~\ref{fig:mceff} shows the acceptance times reconstruction efficiency, 
$\epsilon_{\rm acc}\cdot\epsilon_{\rm reco}$, obtained from the ratio of reconstructed to input $\Lambda$ candidates 
as a function of the proper decay length $L/(\beta\gamma)$, where $L$ denotes the decay length and $\beta$ and $\gamma$ 
are the relativistic velocity and Lorentz factors, respectively. The solid blue curve shows 
$\epsilon_{\mathrm{acc}}\,\cdot\,\epsilon_{\mathrm{reco}}$  for a pure data sample of single $\Lambda$ baryon decays, 
while the dotted line results from $\Lambda$ baryon decays embedded into central PHQMD events.
Both $\epsilon_{\mathrm{acc}}\,\cdot\,\epsilon_{\mathrm{reco}}$ distributions were evaluated 
within the acceptance of $y_{\rm lab}\,\ge\,0.2$.

Both the pure $\Lambda$ sample and the PHQMD events with embedded $\Lambda$ baryon decays were processed through the same 
reconstruction chain, using identical $p_{\rm T}$ and rapidity input distributions. The two simulation setups differ solely 
by the inclusion of additional detector hits in the PHQMD events, arising from particles produced in nucleus–nucleus collisions 
and emitted into the mCBM acceptance. The ratio between the two data samples is displayed in the lower panel 
of Fig.~\ref{fig:mceff}. It is almost independent of the proper decay length, but shows a drop of about $16\%$.
This reduction likely originates from an enhanced fake-track probability in the $\Lambda$-embedded PHQMD sample, driven 
by the minimal hit requirement for track candidates -- two STS hits and one TOF hit -- and the rejection 
of multiple use of detector hits across different candidates. However, since the lifetime extraction only depends 
on the slope of the corrected proper decay length distribution, rather than its absolute normalization, 
the pure $\Lambda$ sample is used for the lifetime analysis because it provides higher statistical precision. 
For the multiplicity measurement, the embedded sample is used instead, as it incorporates the embedding procedure 
and therefore provides a more accurate description of efficiency losses due to background.

The integral geometrical acceptance for $\Lambda \rightarrow p + \pi^-$ decays in the described mCBM setup is 
$\epsilon_{\mathrm{acc}} = (6.5 \pm 0.7) \times 10^{-3}$,
with both daughter particles required to have at least two STS hits and one TOF hit within the detector acceptance.
Applying the selection criteria described in
Sec.~\ref{ssec:reco}, the reconstruction efficiency amounts to
$\epsilon_{\mathrm{reco}} = (4.6 \pm 0.9) \cdot 10^{-2}$. 

\section{Reconstructed $\Lambda$ baryons} 
\label{sec:results}

\subsection{Comparisons with simulation} 
\begin{figure}[pos=ht] 
\centering
\includegraphics[width=.60\linewidth]{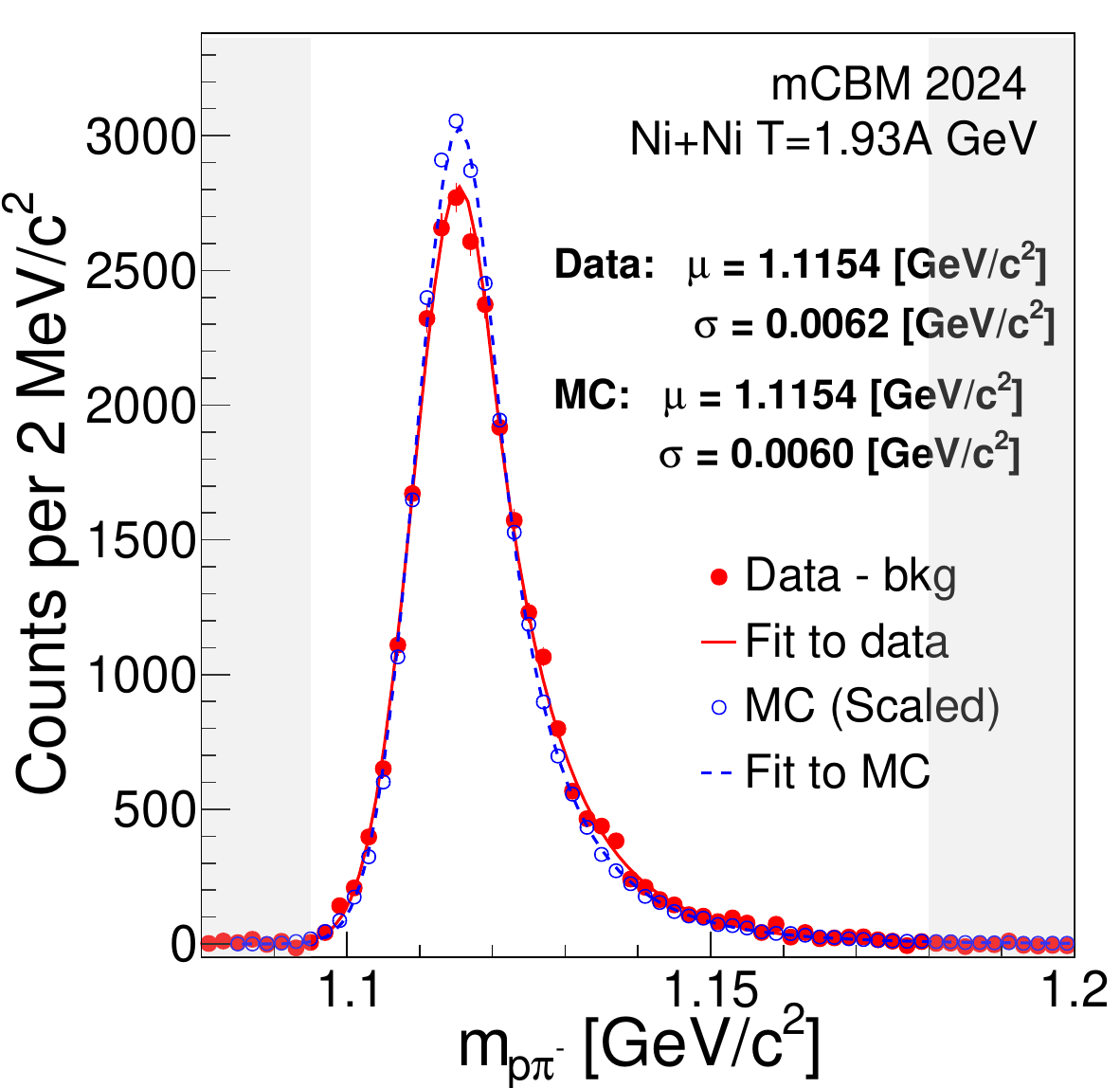} 
\caption{Invariant mass distribution comparing data (red dots) and $\Lambda$ MC simulation (black dots), 
with the MC normalized to match the total data counts. Fits using the Crystal Ball function are represented 
by red and blue solid lines for data and simulation. The gray-shaded areas indicate the invariant-mass regions outside 
the selected range of \SIrange{1.095}{1.180}{GeV/c^2}.}
\label{fig:data_mc_signal}
\end{figure}
    
The background-subtracted invariant mass distribution is compared to simulations in Fig.~\ref{fig:data_mc_signal}. 
For a direct comparison of the spectral shape, the simulated distribution is normalized to the same total yield as the data. 
Similar to the data, the simulation also exhibits an asymmetric distribution. To quantify the peak position and resolution, 
Crystal Ball functions~\cite{PhysRevD.34.711} were fitted to the invariant mass spectra. The fitted mean values in data 
and simulation agree with each other. However, both values are shifted relative to the nominal 
$\Lambda$ mass~\cite{ParticleDataGroup:2024cfk} by approximately $0.3$~MeV/$c^2$. Such a shift may arise from small remaining biases 
in the reconstruction procedure. The fits further indicate that the width observed in data is slightly larger than that 
in the simulation. Two effects may contribute to this difference:  
(i) an overestimation of the timing resolution of the BMON and TOF systems in the simulation;
(ii) residual misalignment of the STS in the data, resulting in additional smearing 
of the reconstructed invariant mass distribution.
 
\begin{figure}[pos=p] 
\centering
\includegraphics[width=.37\linewidth]{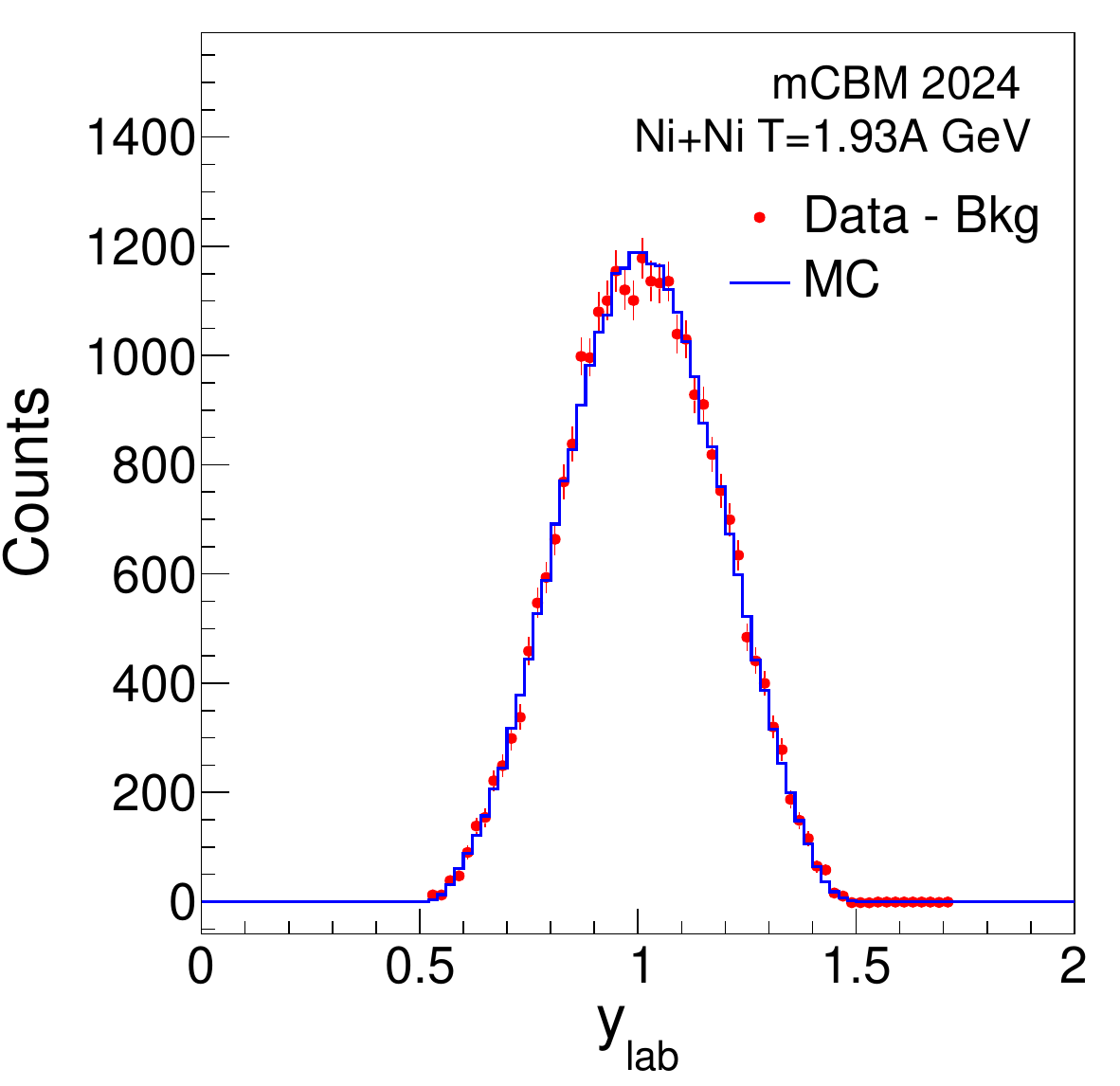} 
\includegraphics[width=.37\linewidth]{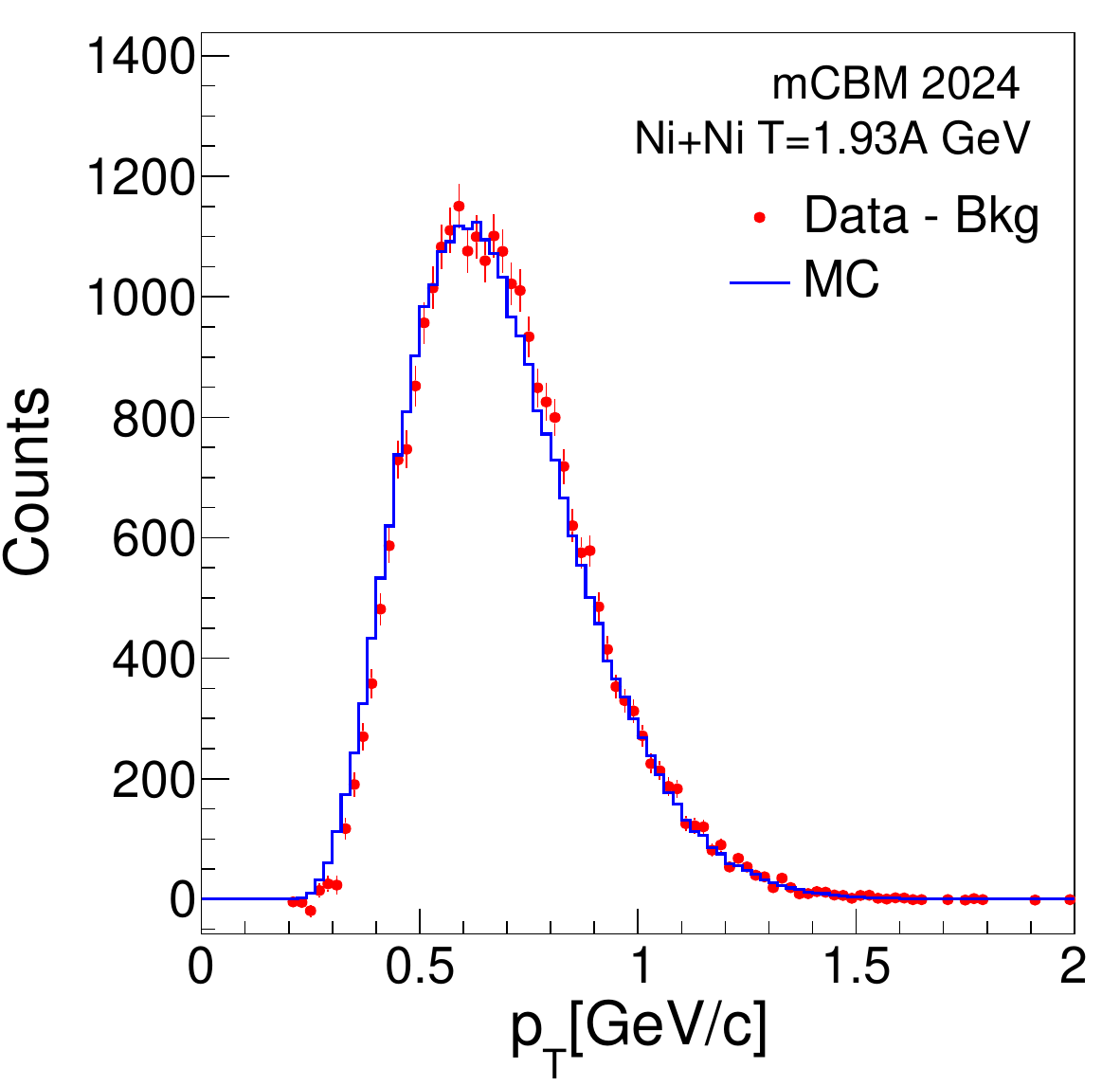} 
\caption{Rapidity $y_{\rm{lab}}$ and transverse momentum $p_{T}$ distributions of $\Lambda$ candidates 
in the invariant mass range \mbox{\SIrange{1.095}{1.180}{GeV/c^2}}. Red dots represent the background-subtracted data, 
while the blue lines represent the simulation. }
\label{fig:lambda_kinematics}
\vspace{5mm}
\includegraphics[width=.37\linewidth]{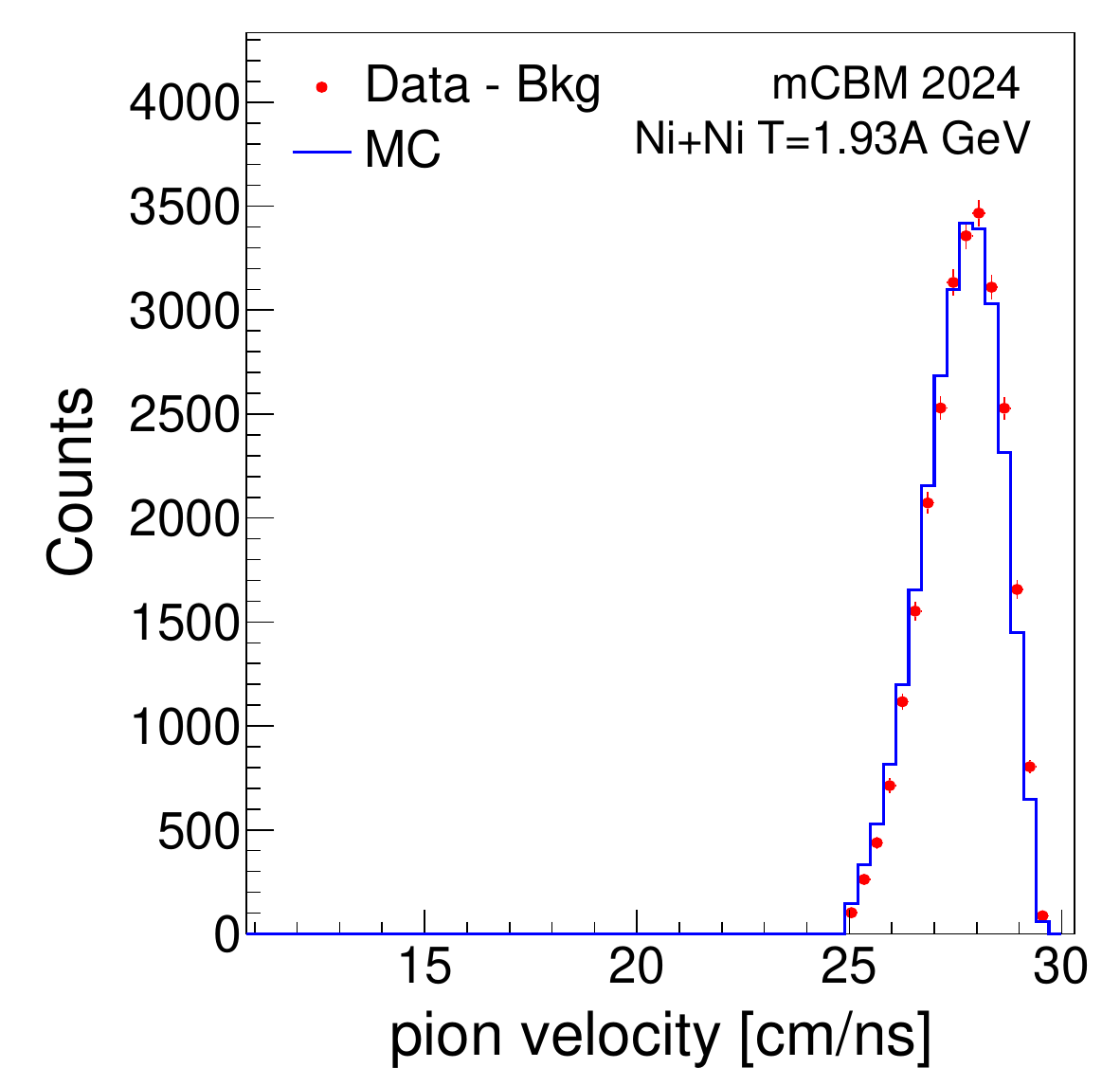} 
\includegraphics[width=.37\linewidth]{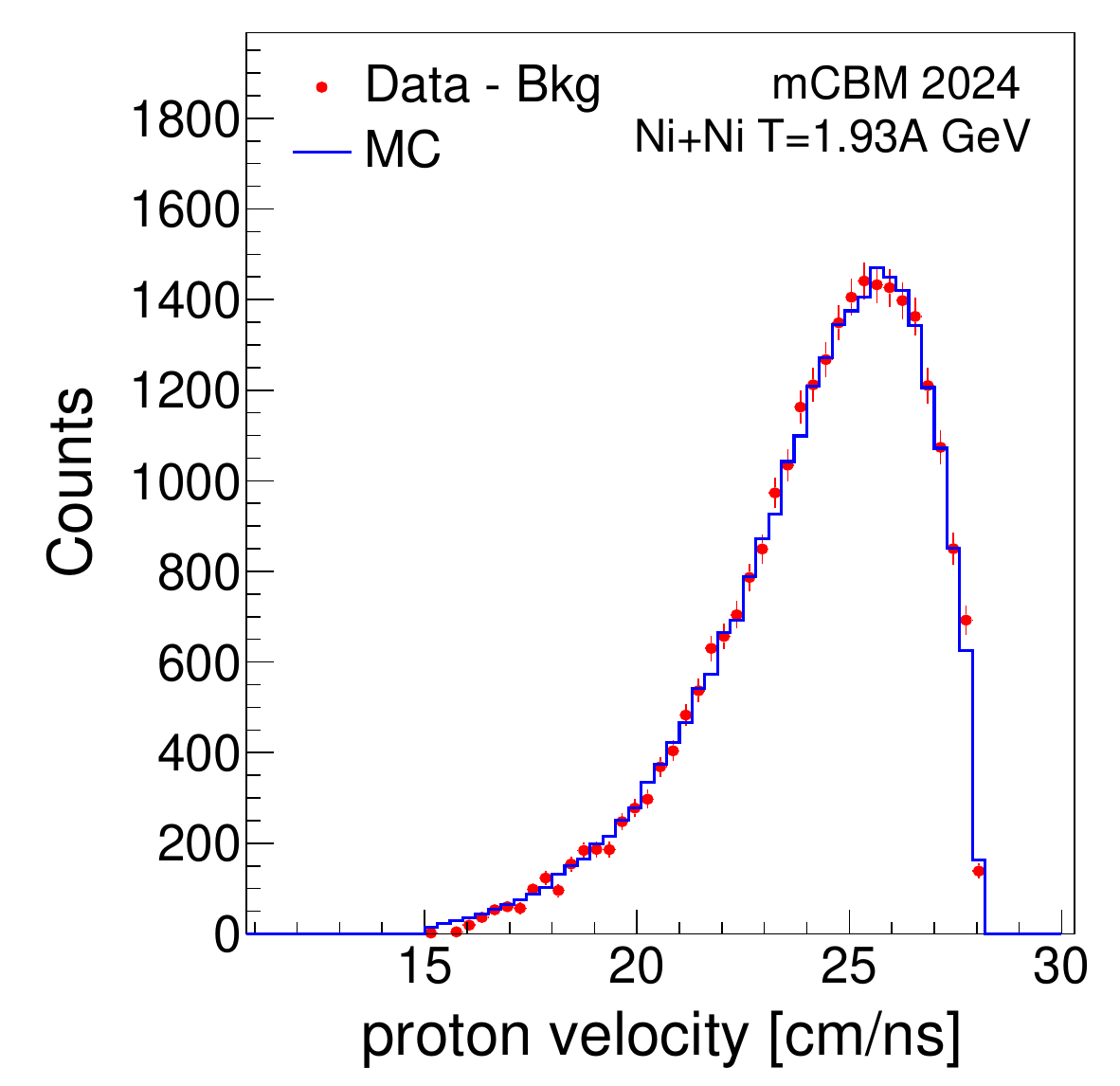} 
\caption{Daughter velocity distributions of $\Lambda$ candidates in the invariant mass range \SIrange{1.095}{1.180}{GeV/c^2}. 
Red dots represent the background-subtracted data, while the blue lines represent the simulation. }
\label{fig:daughter_velocity}
\vspace{5mm}
\includegraphics[width=.245\linewidth]{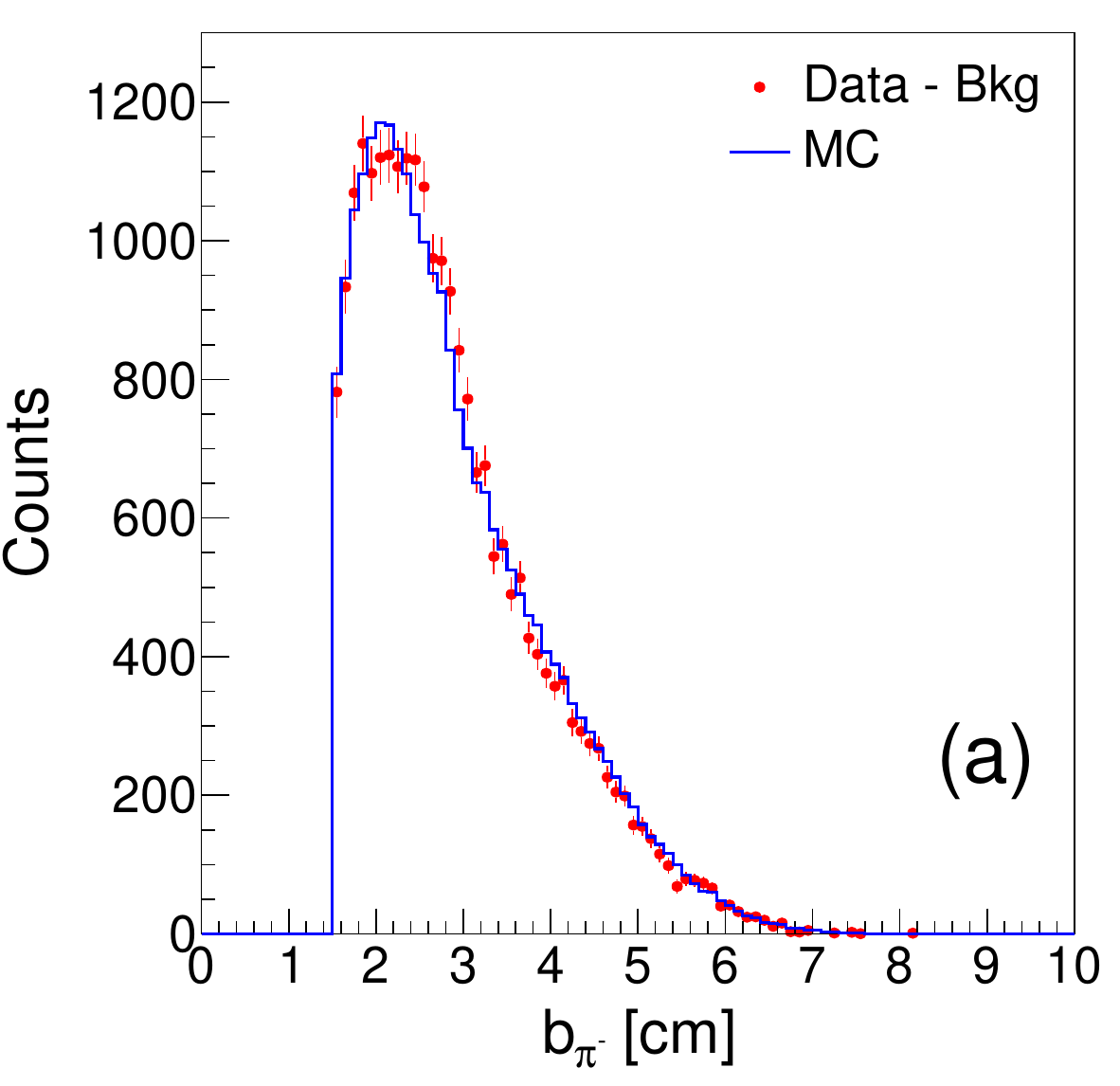} 
\includegraphics[width=.245\linewidth]{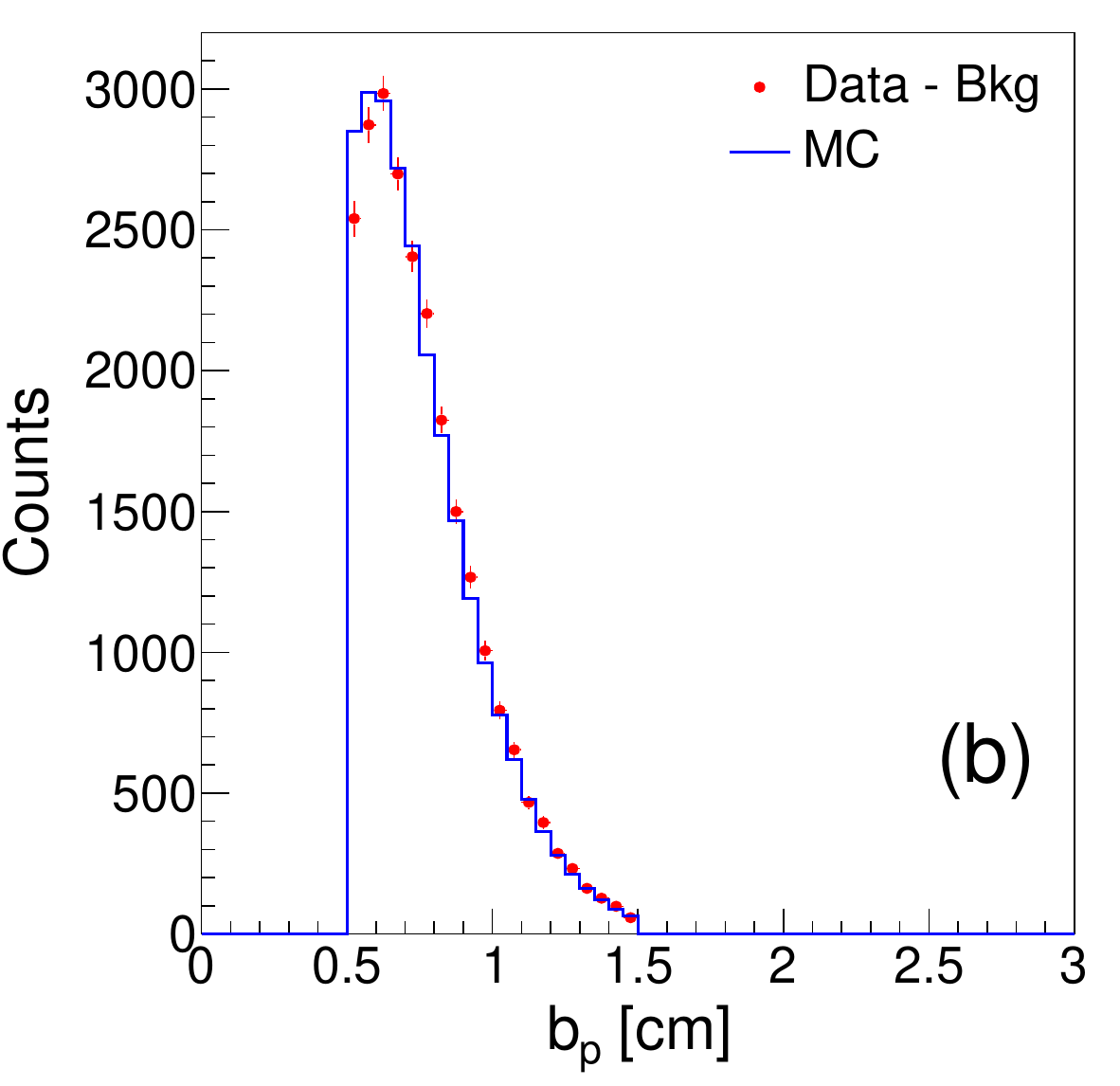} 
\includegraphics[width=.245\linewidth]{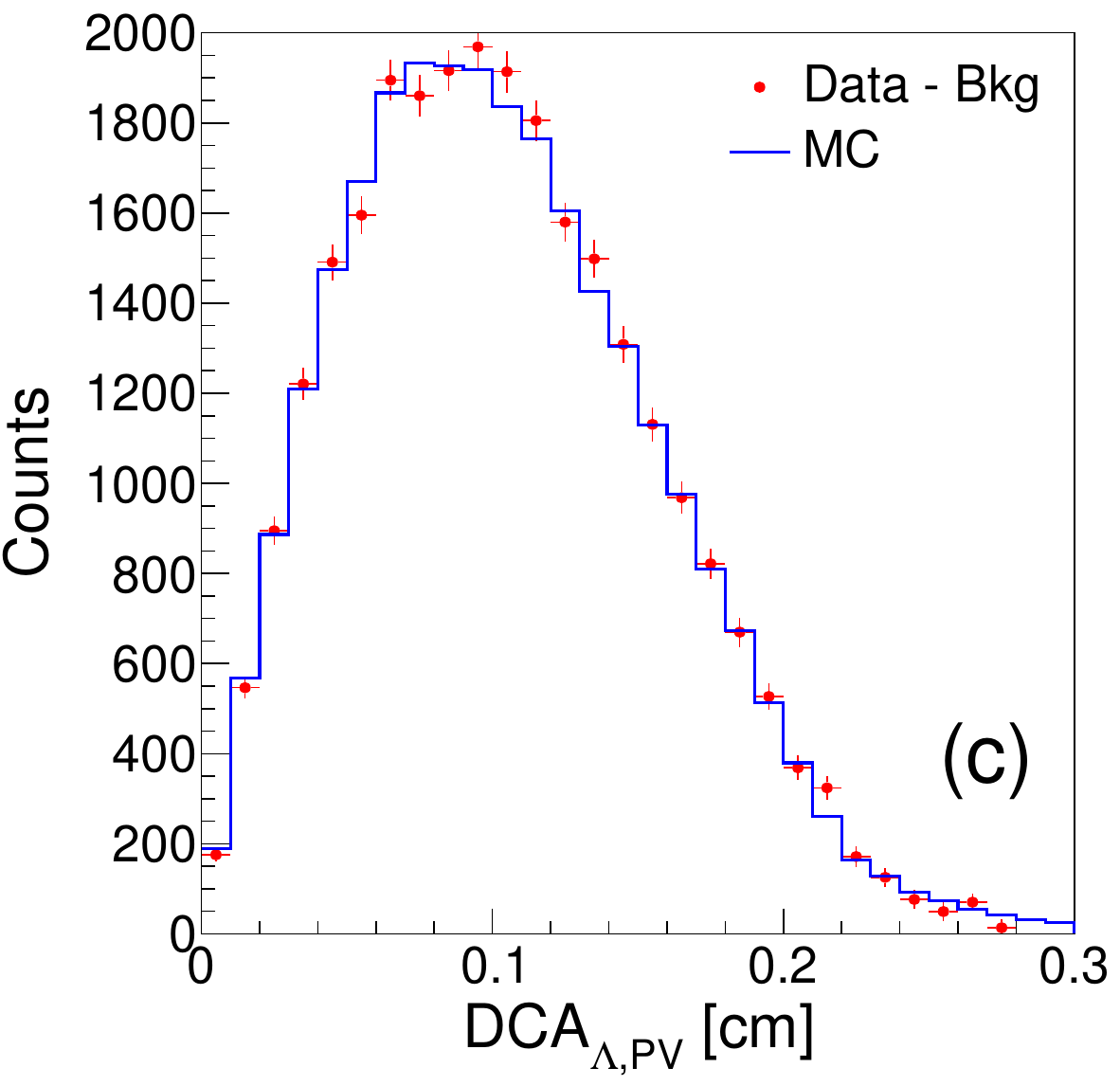} 
\includegraphics[width=.245\linewidth]{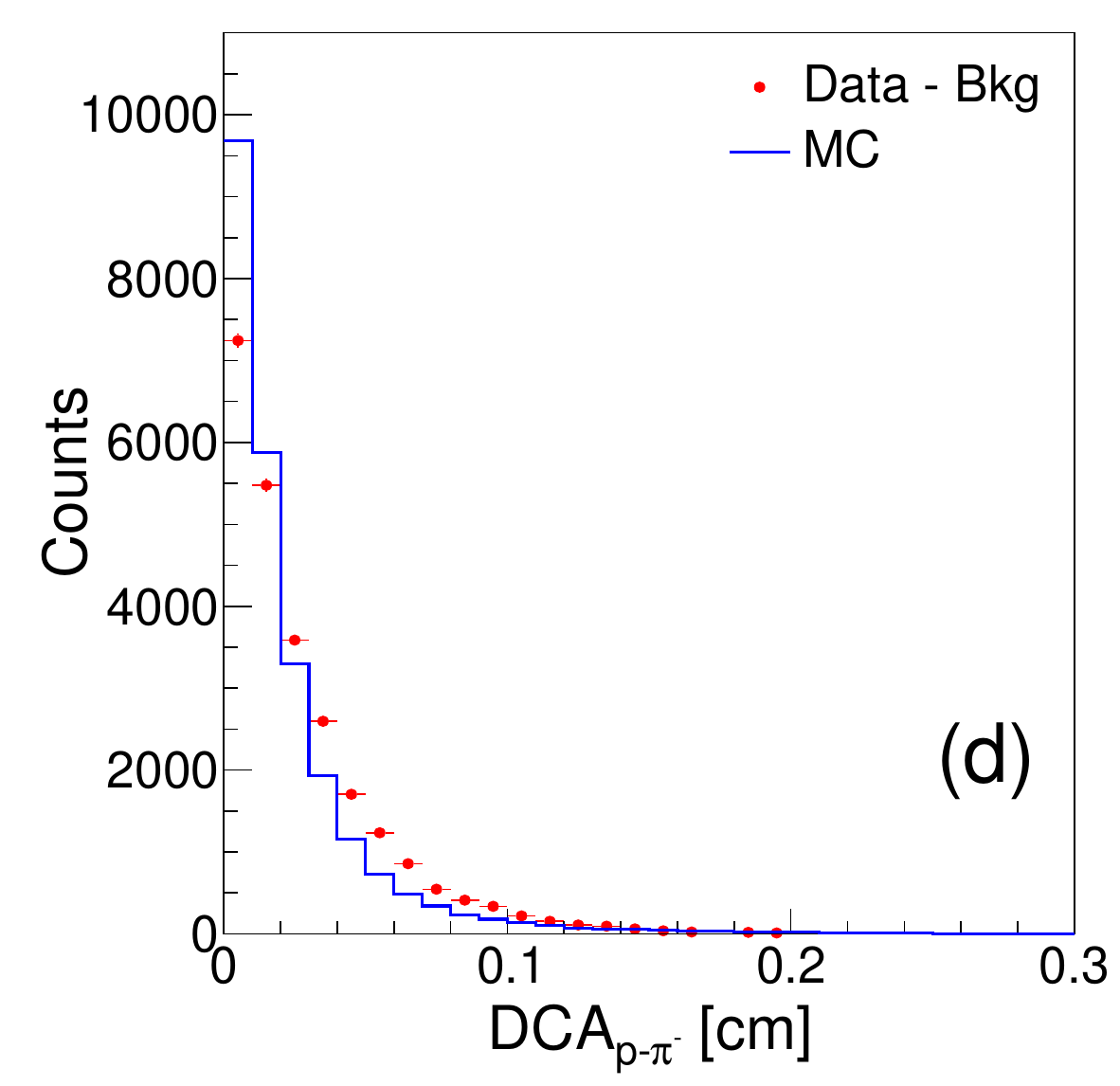} 
\caption{Topological variables of $\Lambda$ candidates in the invariant mass range \SIrange{1.095}{1.180}{GeV/c^2}: 
(a) transverse impact parameter of the pion $b_{\pi^{-}}$, (b) transverse impact parameter of the proton $b_{p}$, 
(c) DCA between the $\Lambda$ candidate and the primary vertex $DCA_{\Lambda-PV}$, and 
(d) DCA between the daughter proton and pion $DCA_{p-\pi^{-}}$. Red dots represent the background-subtracted data, 
blue lines the simulation. }
\label{fig:dca}
\end{figure}

To further validate that the observed signal originates from $\Lambda$ decays, various observables are compared between 
background-subtracted data (red dots) and simulation (blue lines) within the invariant mass range 
of \SIrange{1.095}{1.180}{GeV/c^2}. This mass range captures $>99\%$ of the excess. The distributions are essentially 
reproduced by the simulation, as is visible in Fig.~\ref{fig:lambda_kinematics}~-~\ref{fig:dca}. 
Fig.~\ref{fig:lambda_kinematics} shows the rapidity $y_{\rm lab}$ and transverse-momentum $p_{T}$ distributions. 
As can be seen, the agreement between data and simulation is good, indicating that our choice of the $\Lambda$ source parameters 
($T_{\rm eff}$ and $\sigma_y$) is reasonable. Fig.~\ref{fig:daughter_velocity} presents the velocity distributions 
of the daughter particles. Once again, the simulations accurately reproduce the data. Finally, Fig.~\ref{fig:dca} 
displays various topological distributions of the $\Lambda$ candidates. These include the transverse impact parameter 
of the pion $b_{\pi^{-}}$, the transverse impact parameter of the proton $b_{p}$, the DCA between the $\Lambda$ candidate 
and the primary vertex $DCA_{\Lambda-PV}$, and the DCA between the daughter proton and pion $DCA_{p-\pi^{-}}$. As expected, 
the $b_{p}$ distribution is significantly narrower than the $b_{\pi^{-}}$ distribution due to the larger mass of the proton. 
The $DCA_{\Lambda-PV}$ distribution peaks near zero, which is consistent with $\Lambda$ particles being primarily produced 
at the collision vertex. 
The width reflects contributions from the beam profile and detector resolution. As for $DCA_{p-\pi^{-}}$, the simulation captures 
the overall shape, while the distribution in data is slightly broader, likely due to a residual misalignment of the STS detectors. 
Taken together, these comparisons demonstrate that the excess of the data over the combinatorial background can be attributed 
to real $\Lambda \rightarrow p + \pi^{-}$ decays, and that the simulation provides a realistic and reliable description 
of the $\Lambda$ production and decay kinematics. 

\subsection{$\Lambda$ Lifetime} 
\label{ssec:lifetime}
\begin{figure}[pos=ht] 
\centering
\includegraphics[width=.6\linewidth]{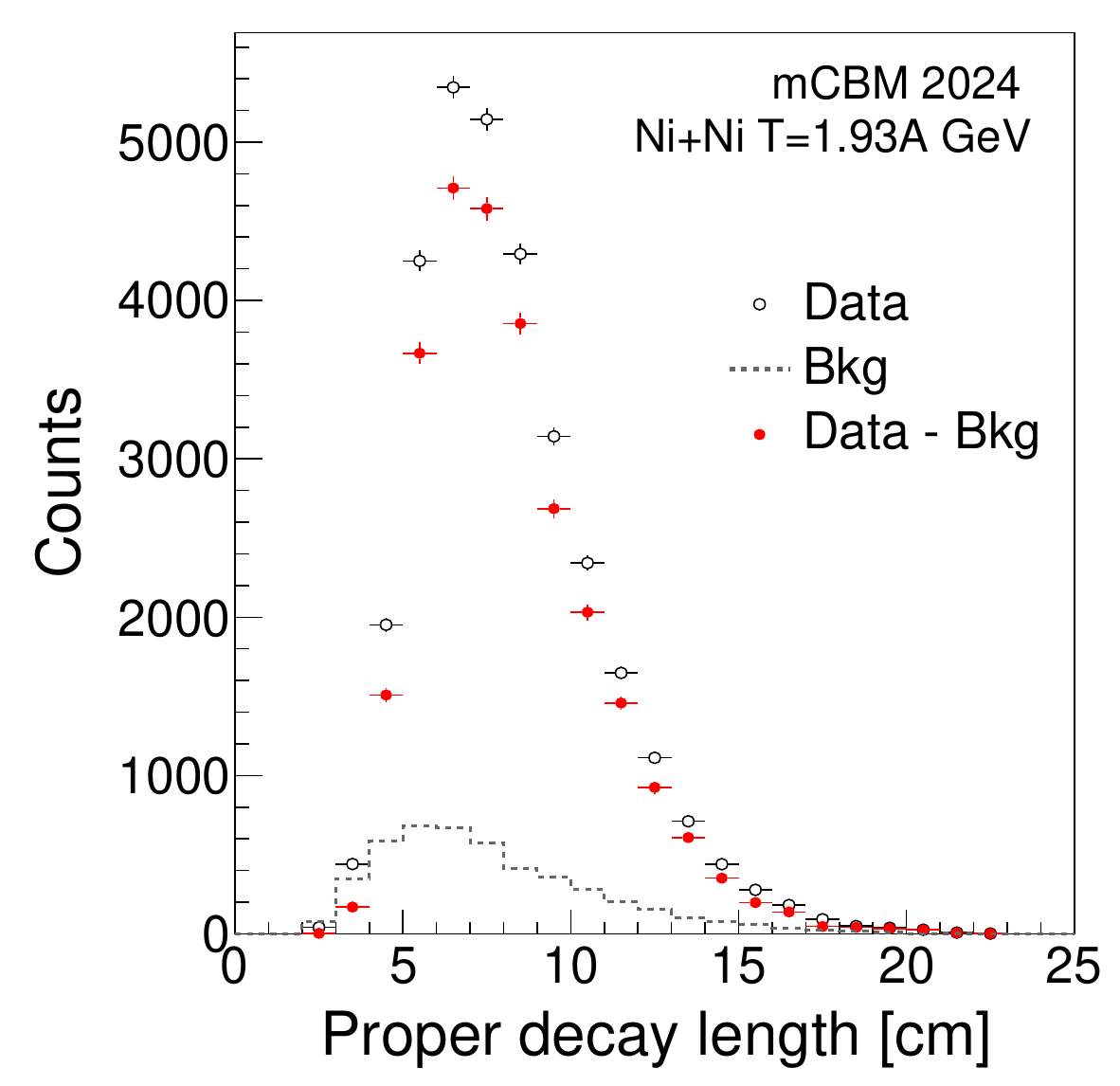}                 
\caption{Proper decay length distributions from the data (open black dots), the combinatorial background (dotted gray line), 
and the background-subtracted data (solid red dots). }
\label{fig:lambda_pl}
\end{figure}

Experimentally, the proper decay length $L/\beta\gamma$ is calculated as $Lm/p$, where $L$ and the momentum $p$ is obtained 
from KFParticle, and $m$ is the mass of the $\Lambda$, \SI{1.115683}{GeV/c^2}~\cite{ParticleDataGroup:2024cfk}. The raw yield in 
various $L/\beta\gamma$ bins is extracted using the method described in Sec.~\ref{sec:signalextraction} and is presented 
in Fig.~\ref{fig:lambda_pl}.

\begin{figure}[pos=ht]
\centering
\includegraphics[width=.65\linewidth]{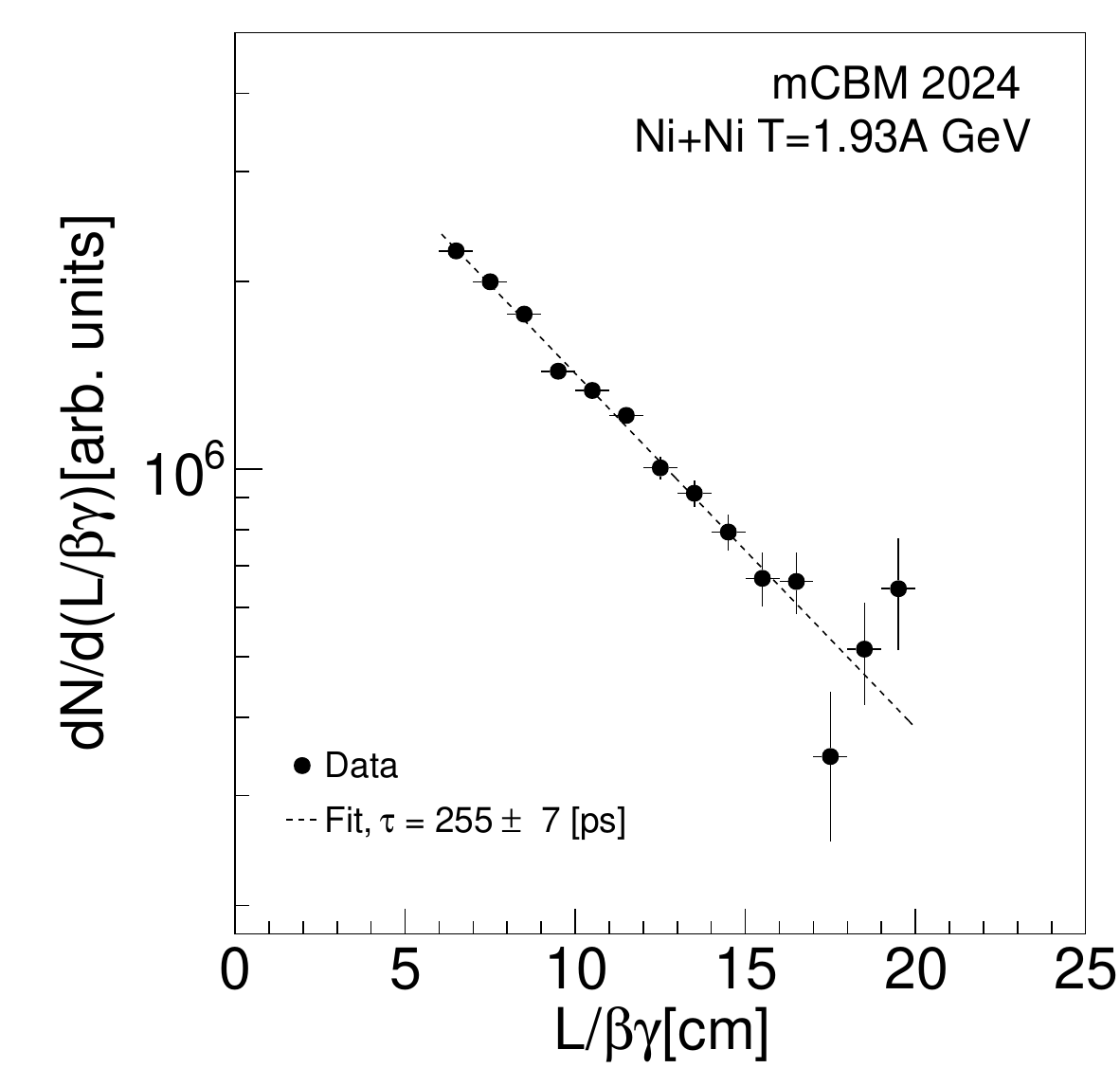}
\caption{Acceptance and efficiency corrected proper decay length distributions from Ni+Ni collisions at $T=1.93$ AGeV.
The dashed line represents an exponential fit.}
\label{fig:lambda_lifetime}
\end{figure}

\noindent
The acceptance and efficiency corrected proper decay length distributions can be obtained using the relation:
\begin{equation}
\frac{dN}{d(L/\beta\gamma)} = \frac{1}{\epsilon_{\rm acc}\cdot\epsilon_{\rm reco}(L/\beta\gamma)} 
    \frac{\Delta N_{\text{sig}}(L/\beta\gamma)}{\Delta (L/\beta\gamma)} ~~~,
\end{equation} 
where $\Delta N_{\text{sig}}(L/\beta\gamma)$ is the number of signal counts in a $L/\beta\gamma$ bin, $\Delta (L/\beta\gamma)$ 
is the width of the $L/\beta\gamma$ bin, and $\epsilon_{\rm acc}\cdot\epsilon_{\rm reco}$ is the acceptance and efficiency 
determined in Sec.~\ref{sec:eff}. The final acceptance and efficiency corrected proper decay length distributions are shown 
in Fig.~\ref{fig:lambda_lifetime}.

If the excess in the data originates from real $\Lambda$ decays, the proper decay length distribution should follow an exponential 
decay law:
\begin{equation}
N(L/\beta\gamma) = N_{0} e^{-L / (\beta\gamma c\tau)} ~~~ ,
\end{equation}
with the $\Lambda$ lifetime indicated by $\tau$. The fit to the background subtracted data  yields $\tau = 255 \pm 7$\,(stat.) ps.

\subsubsection{Systematic uncertainties}         
To evaluate the robustness of our lifetime measurement, we assess its systematic uncertainties. These represent potential deviations 
arising from various aspects of the analysis and the simulation. The following sources of systematic uncertainty are considered.

\paragraph{\textbf{Signal extraction}} 
The reconstructed $\Lambda$ signal exhibits an asymmetric invariant mass distribution due to the reconstruction method. 
While the left side of the peak is well described by a Gaussian, the right side features a pronounced tail, complicating 
the separation between signal and background in that region. To estimate the associated systematic uncertainty, 
three aspects of the signal extraction procedure are varied. 
First, two different normalization schemes for the mixed-event background are tested:  
(i) normalizing the mixed events to the data in the sideband region using the full dataset, and applying 
the same normalization factor across all proper lifetime bins;  
(ii) normalizing the mixed events independently for each proper lifetime bin.  
The average of the two results is taken as the default value, and half of their difference, which amounts to \SI{7}{ps}, 
is assigned as the systematic uncertainty from this source. Second, the invariant mass window used to define 
the signal region is reduced to \SIrange{1.095}{1.140}{GeV/c^2}, resulting in a change of $+11$~ps in the extracted lifetime. 
Third, the invariant mass range used for background normalization is modified to \SIrange{1.083}{1.090}{GeV/c^2}, 
leading to a change of $<$\SI{1}{ps} in the extracted lifetime. The total systematic uncertainty from these variations 
is obtained by summing the individual contributions in quadrature, yielding a combined uncertainty of \SI{13}{ps}.

\paragraph{\textbf{Tracking efficiency}}     
The mismodeling of tracking efficiency in Monte Carlo simulations can lead to an incorrect estimation 
of the true tracking efficiency, which could bias the measured lifetime. Contributing factors 
include residual detector misalignments in data, hit efficiency in simulations, etc. To estimate the systematic uncertainty
arising from these effects, two variation studies were performed.  
\begin{enumerate}
    \item STS hit efficiency: The efficiency of the STS station STS2 --~where data indicate the presence 
    of inefficient channels~-- was varied from its nominal value of 80\% to 70\% and 90\%. Both resulted 
    in approximately \SI{6}{ps} change in the fitted lifetime, which was assigned as the uncertainty.  
    \item Detector misalignment: The track residuals in various STS sensors are observed to be 
    on the order of \SI{100}{\micro m}. Thus, misalignment tolerances of $\pm$ \SI{200}{\micro m} and $\pm$ \SI{400}{\micro m} 
    were introduced by artificially inflating the hit position uncertainties in the tracking 
    algorithm. The fitted lifetimes for these scenarios were $246 \pm 6$\,ps and $264 \pm 6$\,ps, 
    respectively, yielding a root-mean-square (RMS) corresponding to a \SI{9}{ps} uncertainty.
\end{enumerate}
Adding these two contributions in quadrature, we assign a total systematic uncertainty from tracking efficiency 
mismodeling of \SI{11}{ps}.

\paragraph{\textbf{Topological variable modeling}} 
The description of topological variables in the Monte Carlo simulation including the (i) DCA between the daughter tracks, 
(ii) the DCA between the daughter pion/proton track and (iii) the primary collision vertex, and (iv) the DCA between 
the reconstructed mother particle and the collision vertex may not fully reflect those observed in experiment data. 
This discrepancy arises from factors such as residual detector misalignment, inaccuracies in the material budget 
(which affect energy loss and multiple scattering), and imperfect modeling of detector hit efficiency, which influences 
the pointing resolution.

To estimate the impact of these effects, the selection criteria applied to all four aforementioned topological variables 
were varied by $\sim20\%$ of the default cut value. The total systematic uncertainty is obtained by summing 
the differences in the fitted lifetime to the default value in quadrature, yielding to an overall uncertainty of \SI{16}{ps}.

\paragraph{\textbf{Input kinematic distributions}} 
For simulations, the input rapidity distribution is modeled using a Gaussian function, and the transverse momentum distribution 
is described by a Boltzmann function. The width of the Gaussian and the effective temperature are derived from FOPI measurements 
as described in Section~\ref{ssec:lambdaparam}. To estimate the associated systematic uncertainty, the Gaussian width 
and the Boltzmann temperature were varied independently by 1$\sigma$ of the quoted uncertainty. The $\pm$1$\sigma$ variations 
in the parameter values for the rapidity width and the effective temperature result in an average \SI{15}{ps} change 
in the extracted lifetime respectively, which is assigned as systematic uncertainty.

\paragraph{\textbf{Total systematic uncertainties}}
A summary of the systematic uncertainties of the $\Lambda$ lifetime extraction can be found in Tab.~\ref{tab:systematics}. 
Adding all sources in quadrature, the total systematic uncertainty is \SI{28}{ps}. 

\subsubsection{Comparison with published results}
The reconstructed lifetime yields to 
\mbox{$\tau = (255 \pm 7 ~\rm{(stat.)} \pm 28 ~\rm{(syst.)}$)\,\si{ps}}, 
which is in agreement with the Particle Data Group value of (263\,$\pm$\,2)\,\si{ps} and provides evidence 
that the observed excess corresponds to genuine $\Lambda$ decays. 

\begin{table}[t]
    \centering
    \caption{Systematic uncertainty sources for the $\Lambda$ lifetime extraction.}
    \label{tab:systematics}
    \begin{tabular}{@{}lS[table-format=1.1]@{}}
        \toprule
        \textbf{Source} & \textbf{Uncertainty} \\
        \midrule
        Signal extraction                           & \SI{13}{ps} \\
        Tracking efficiency                         & \SI{11}{ps} \\
        Topological variable modeling               & \SI{16}{ps} \\
        Input kinematic distributions               & \SI{15}{ps} \\
        \midrule
        \textbf{Total}                              & \textbf{\SI{28}{ps}} \\
        \bottomrule
    \end{tabular}
\end{table}

\subsection{$\Lambda$ Multiplicity} 
\label{ssec:YieldEstimate} 
Using the MC simulation, the geometrical acceptance was determined to be 
$\epsilon_{\mathrm{acc}} = (6.5\,\pm\,0.7) \cdot 10^{-3}$ and the reconstruction efficiency to be 
$\epsilon_{\mathrm{reco}} = (4.6\,\pm\,0.9) \cdot 10^{-2}$, 
see Sec.~\ref{sec:sim}. Accordingly, the $\Lambda$ multiplicity results to
\begin{equation}
M_{\Lambda, \rm{trig}} = 
\frac{ N_{\Lambda}^{\mathrm{reco}} \cdot \epsilon_{\mathrm{acc}}^{-1} \cdot \epsilon_{\mathrm{reco}}^{-1} } { N_{\mathrm{event}} } 
= 0.031 \pm 0.0002 ~ \rm{(stat.)}
\label{eq:M_Lambda}
\end{equation}
for the triggered sample of Ni\,+\,Ni collisions at 1.93\,AGeV kinetic projectile energy ($\sqrt{s_{\rm{NN}}}=2.67$\,GeV). 

To enable a comparison with the published results of the FOPI Collaboration~\cite{FOPI:2007usx}, the measured $\Lambda$ multiplicity 
in the triggered event sample is scaled to central Ni+Ni collisions. The triggered sample corresponds to a sampled cross-section of $\sigma_{\rm trigger} = (2.2 \pm 0.2)\,\rm{b}$\,, 
while central collisions in FOPI correspond to $\sigma_{\rm cen} = \SI{0.35}{b}$. A central-collision scaling factor of 
$f_{\rm cen}=2.94$ is obtained from PHQMD Monte Carlo simulations (see Section~\ref{ssec:phqmd}). The scaling factor is consistent with $\Lambda$ multiplicity measurements at comparable beam energies~\cite{STAR:2024znc,HADES:2018noy}. Applying this scaling factor to the observed $\Lambda$ multiplicity~(\ref{eq:M_Lambda}) results in 
\begin{equation}
M_{\Lambda, \rm{cen}} = f_{\rm cen}\cdot M_{\Lambda,\rm{trig}}
= 0.091 \pm 0.0006 ~ \rm{(stat.)}.
\end{equation}

\subsubsection{Systematic Uncertainties}
Similar to the lifetime analysis, we evaluate various sources of systematic uncertainty. 

\paragraph{\textbf{Signal extraction}} The systematic uncertainties 
associated with signal extraction are assessed 
using the same procedures as in the lifetime study.

\paragraph{\textbf{Tracking efficiency}} For the tracking efficiency, in addition to the procedures that we carried out for the lifetime study to assess 
the systematic uncertainty, we also consider the systematic uncertainty related to the hit reconstruction efficiency 
and misalignment. These contributions were negligible in the lifetime analysis, since the hit reconstruction efficiency 
and misalignment effects are not expected to exhibit a significant dependence on the particle decay length. 
For the yield measurement, however, their impact is included. The uncertainty in the hit reconstruction efficiency 
is estimated to be $12\%$ based on detector performance studies~\cite{CBM:2025voh}, while the uncertainty due to residual misalignment 
is estimated to be $12\%$ by varying the hit positions within their estimated uncertainties in the simulation. The total uncertainty 
in the tracking efficiency is estimated to be $19\%$ by summing these components in quadrature. 

\paragraph{\textbf{Topological variable modeling}}
The systematic uncertainties 
associated with topological variable modeling are assessed 
using the same procedures as in the lifetime study.

\paragraph{\textbf{Phase space extrapolation}} For the yield measurement, the systematic uncertainty associated with the
extrapolation to full phase space is evaluated in two steps. First, the uncertainty of the transverse momentum extrapolation 
is obtained comparing the results by employing Boltzmann and blast-wave parameterizations. In a second step, this uncertainty 
is propagated to the rapidity extrapolation by performing Gaussian fits within the corresponding $p_{T}$ extrapolation
uncertainties. The resulting variation of the extrapolated $4\pi$ yield is $17\%$, which is assigned as the 
systematic uncertainty associated with the extrapolation to the unmeasured kinematic region. 

\paragraph{\textbf{Central-collision scaling factor}} An uncertainty of 0.30 is assigned on the central-collision
scaling factor $2.94$, 
reflecting the variation observed under different PHQMD event settings. 

\paragraph{\textbf{Total systematic uncertainties}} A breakdown of the systematic uncertainties is shown in
Tab.~\ref{tab:systematics_yield}. Adding all sources in quadrature, the total relative systematic uncertainty is $28\%$.

\begin{table}[t]
    \centering
    \caption{Relative systematic uncertainty sources for the central-collision $\Lambda$ multiplicity measurement.}
    \label{tab:systematics_yield}
    \begin{tabular}{@{}lS[table-format=1.1]@{}}
        \toprule
        \textbf{Source} & \textbf{Uncertainty} \\
        \midrule
        Signal extraction                           & \SI{2}{\%} \\
        Tracking efficiency                         & \SI{19}{\%} \\        
        Topological variable modeling               & \SI{3}{\%} \\
        Phase space extrapolation                   & \SI{17}{\%} \\
        Central-collision scaling factor            & \SI{10}{\%} \\
        \midrule
        \textbf{Total}                              & \textbf{\SI{28}{\%}} \\
        \bottomrule
    \end{tabular}
\end{table}

\subsubsection{Comparison with published results}

The estimated $\Lambda$ multiplicity in central collisions is $M_{\Lambda,\rm{cen}}= 0.091 \pm 0.0006 ~ \rm{(stat.)} \pm 0.025 ~ \rm{(syst.)}$ 
and agrees with the FOPI result 
$0.137 \pm 0.005 ~ \rm{(stat.)} ~ ^{+0.009}_{-0.025} ~ \rm{(syst.)}$ 
within the statistical and systematic uncertainties.
  
\section{Conclusions and Outlook} 
\label{sec:summary}
In conclusion, the mCBM experimental setup, which serves as a demonstrator and testbed for the CBM experiment at FAIR, 
was successfully operated at the SIS18 accelerator of GSI.
Prototype and pre-series modules of all CBM subsystems were installed, commissioned, and operated together 
with a free-streaming data-acquisition system under realistic experimental conditions.
This work demonstrates the functionality of essential elements of the CBM experiment, including detector integration, 
continuous data readout, data transport, and the reconstruction of physics observables of rare probes 
using the CBM data-acquisition and reconstruction framework.

As a benchmark for the validation of the detector designs, data-processing chain, and reconstruction algorithms to be used in CBM, 
the reconstruction of $\Lambda \rightarrow p + \pi^{-}$ decays was demonstrated using a small data sample corresponding to 
approximately 5.5~hours of beam on target in Ni+Ni collisions at the beam kinetic energy $T = 1.93$\,AGeV. 
The reconstruction of $\Lambda$ baryons 
serves as a stringent test case, as it combines low production probability with non-trivial background suppression requirements. 
The successful extraction of a statistically significant signal, together with the agreement of topological and kinematic 
distributions with GEANT simulations and a $\Lambda$ lifetime and multiplicity consistent with published values, 
demonstrates that non-trivial and rare observables can be reliably reconstructed from continuous, triggerless detector data 
using the CBM data acquisition and reconstruction framework.

While the $\Lambda$ benchmark demonstrates the viability of the CBM data-acquisition and reconstruction framework, 
several aspects of the present study are specific to the mCBM demonstrator and must be interpreted in the context 
of its reduced scale. The mCBM setup is operated without a magnetic field, such that no momentum determination 
is available and kinematic quantities are derived from time-of-flight information only.
Nevertheless, the successful reconstruction of $\Lambda$ decays under these conditions demonstrates that 
non-trivial decay topologies can be identified using the CBM tracking and reconstruction algorithms even 
in a field-free environment. In the full CBM experiment, the presence of a magnetic field and momentum measurement 
will provide improved tracking performance and momentum resolution, which is expected to further enhance 
signal reconstruction and background suppression. The present $\Lambda$ benchmark analysis is based on data recorded 
at an average interaction rate of about \SI{250}{kHz} while the hit rate on the beam counters of about \SI{5}{MHz} 
is already close to the operational limits of the BMON system of CBM which is planned to reach \SI{10}{MHz} hit rate. 

Looking ahead, the mCBM campaigns from 2022 to 2025 have produced a large body of data for both reconstruction 
of $\Lambda$-baryons and detector studies, including dedicated rate scans and high-rate runs with various beam species 
reaching average interaction rates of up to \SI{10}{MHz}.
In addition, data sets covering excitation functions and system-size dependencies of $\Lambda$-baryon production 
at lower beam energies of \SI{1.58}{AGeV} (Ni+Ni, Ag+Ag) and \SI{1.23}{AGeV} (Ni+Ni, Ag+Ag, Au+Au) have been recorded.
These data, characterized by reduced production probabilities, increased background levels, and higher occupancies, 
will be used to further optimize reconstruction algorithms, their underlying calibrations, and the Monte Carlo description 
within the CBM software framework. Together with continued high-rate commissioning and data challenges, 
these efforts directly contribute to the preparation of CBM data taking foreseen to start in 2028.
  
\section*{Acknowledgment} 
The CBM Collaboration would like to thank the GSI/FAIR accelerator teams for excellent beam conditions 
within the FAIR Phase-0 program and all GSI/FAIR infrastructure departments
for providing resources and excellent support. This work was supported in part by 
the European Union’s Horizon 2020 research and innovation program EURIZON, 
%
the Bundesministerium f\"ur Forschung, Technologie und Raumfahrt (BMFTR, Germany), 
the GSI Helmholtzzentrum für Schwerionenforschung GmbH (Germany), 
the GSI R\&D Program with the Universities of Darmstadt, Frankfurt, Gie{\ss}en, Heidelberg, M\"unster and Wuppertal (Germany), 
the Deutsche Forschungsgemeinschaft (DFG, Germany), 
the Helmholtz Forschungsakademie Hessen f\"ur FAIR (HFHF, Germany),
the Helmholtz International Centre for FAIR (Germany),
the "Netzwerke 2021", an initiative of the Ministry of Culture and Science of the State of Northrine Westphalia (Germany),
the Frankfurt Institute for Advanced Studies (FIAS, Germany),
%
the FAIR-CZ Project infrastructure program (Czech Republic),
the FAIR-CZ Innovation program (Czech Republic), 
the Facility for Antiproton and Ion Research - participation of the Czech Republic (OPII" MEYS OP VVV and OPIII" MEYS OP VVV JAK program),
%
the National Research, Development and Innovation Office (NKFIH, Hungaria),
the Hungarian OTKA fund,
%
the National Research Foundation of Korea,
%
the Department of Science and Technology (DST, India),
%
the Japan Society for the Promotion of Science, 
%
the Polish Ministry of Education and Science program “Premia na Horyzoncie 2”,
the statutory funds of the Institute of Electronic Systems, Warsaw (Poland),
the Warsaw University of Technology under the Excellence Initiative – Research University (ID-UB) program (Poland),
the National Science Center (Poland), 
%
the Romanian Ministry of Education and Research / National Research Authority,
and the Institute of Atomic Physics (IFA, Romania).
  
\clearpage
  

\printcredits

\bibliographystyle{unsrtnat}

\bibliography{main}



\end{document}